\documentclass[fleqn,usenatbib]{mnras}

\usepackage{newtxtext,newtxmath}

\usepackage[T1]{fontenc}

\DeclareRobustCommand{\VAN}[3]{#2}
\let\VANthebibliography\thebibliography
\def\thebibliography{\DeclareRobustCommand{\VAN}[3]{##3}\VANthebibliography}

\usepackage{graphicx}	
\usepackage{amsmath}	
\usepackage{amssymb}	
\usepackage{color}
\usepackage{mathabx}
\usepackage{epstopdf}

\title[MRI-active inner disc]{MRI-active inner regions of protoplanetary discs. II. Dependence on dust, disc and stellar parameters}

\author[]{Marija R. Jankovic$^{1}$\thanks{E-mail: mj577@cam.ac.uk},
Subhanjoy Mohanty$^{2}$,
James E. Owen$^{2}$
and Jonathan C. Tan$^{3, 4}$
\\
$^{1}$Institute of Astronomy, University of Cambridge, Madingley Road, Cambridge CB3 0HA, UK\\
$^{2}$Astrophysics Group, Imperial College London, Blackett Laboratory, Prince Consort Road, London SW7 2AZ, UK\\
$^{3}$Dept. of Astronomy, University of Virginia, Charlottesville, Virginia 22904, USA\\
$^{4}$Dept. of Space, Earth \& Environment, Chalmers University of Technology, Gothenburg SE-412 96, Sweden
}

\date{Accepted XXX. Received YYY; in original form ZZZ}

\pubyear{2020}

\begin{document}
\label{firstpage}
\pagerange{\pageref{firstpage}--\pageref{lastpage}}
\maketitle

\begin{abstract}

Close-in super-Earths are the most abundant exoplanets known. It has been hypothesized that they form in the inner regions of protoplanetary discs, out of the dust that may accumulate at the boundary between the inner region susceptible to the magneto-rotational instability (MRI) and an MRI-dead zone further out.
In Paper I we presented a model for the viscous inner disc which includes heating due to both irradiation and MRI-driven accretion; thermal and non-thermal ionization; dust opacities; and dust effects on ionization. 
Here we examine how the inner disc structure varies with stellar, disc and dust parameters. For high accretion rates and small dust grains, we find that: (1) the main sources of ionization are thermal ionization and thermionic and ion emission; (2) the disc features a hot, high-viscosity inner region, and a local gas pressure maximum at the outer edge of this region (in line with previous studies); and (3) an increase in the dust-to-gas ratio pushes the pressure maximum outwards. Consequently, dust can accumulate in such inner discs without suppressing the MRI, with the amount of accumulation depending on the viscosity in the MRI-dead regions. Conversely, for low accretion rates and large dust grains, there appears to be an additional steady-state solution in which: (1) stellar X-rays become the main source of ionization; (2) MRI-viscosity is high throughout the disc; and (3) the pressure maximum ceases to exist. Hence, if planets form in the inner disc, larger accretion rates (and thus younger disks) are favoured.

\end{abstract}




\section{Introduction}
Exoplanet discoveries have shown that close-in super-Earths, planets with radii of 1--4\,R$_\oplus$ and orbital periods shorter than $\sim$100 days, are extremely common \citep{Fressin2013, Dressing2013, Dressing2015, Mulders2018, Hsu2019}. To form the solid cores of these planets requires more mass in solids than is expected to exist at short orbital periods in the initial phases of planet formation in protoplanetary discs \citep{Raymond2008, Chiang2013, Schlichting2014, Raymond2014}. Because of this, it has been proposed that these super-Earths form further away from the star, in regions where the temperature in the protoplanetary disc is low enough for water ice to condense, which increases the total amount of solids. In this hypothesis, ice-rich planets migrate inwards, to their present orbits, through gravitational interactions with the disc \citep{Terquem2007, Ogihara2009, McNeil2010, Cossou2014, Izidoro2017, Izidoro2019, Bitsch2019}. However, when compared against atmospheric evolution models, the observed radius distribution of close-in super-Earths is found to be consistent with their cores being rocky, with very little ice present \citep{Owen2017, VanEylen2018, Wu2019,Rogers2021}.

This possibly implies that close-in super-Earths form in the inner, hot regions of protoplanetary discs, near their present orbits. As noted above, it is not expected that these inner regions contain, initially, enough mass in solids to form the super-Earths. However, the inner disc can be enriched by pebbles from the outer disc \citep{Hansen2013, Boley2013, Chatterjee2014, Hu2018, Jankovic2019}, as pebbles are prone to inwards radial drift due to gas drag \citep{Weidenschilling1977}. It has been hypothesized \citep{Chatterjee2014} that the radial drift of pebbles could be stopped at a local gas pressure maximum in the inner disc. Over time the pressure maximum could accumulate enough material to form a super-Earth-sized planet. 

A gas pressure maximum is expected to form in an inner disc that accretes viscously via turbulence induced by the magneto-rotational instability \citep[MRI;][]{Balbus1991, Kretke2009, Dzyurkevich2010, Chatterjee2014}. The susceptibility of the disc to the MRI depends on the coupling between the gas and the magnetic field, and thus on the ionization fraction in the disc. In the hot innermost disc, the MRI is expected to drive high viscosity (i.e., efficient accretion) as a result of thermal ionization. At larger distances, where gas is colder and the ionization fraction drops, the viscosity is expected to be low \citep[such a region is called a dead zone;][]{Gammie1996}. In steady-state, a local gas pressure maximum forms at the transition between the high-viscosity and the low-viscosity regions (the dead zone inner edge) \citep[e.g.][]{Terquem2008}. 


The pressure maximum will only trap pebbles which are prone to radial drift relative to the gas. Smaller dust grains that are well coupled to the gas may be advected and diffused through the pressure maximum by the gas accreting onto the star. In the inner disc, the size of dust grains is limited by fragmentation due to relative turbulent velocities \citep{Birnstiel2010, Birnstiel2012, Drazkowska2016}. Pebbles that radially drift from the outer to the inner disc become smaller due to fragmentation, and the effect of radial drift weakens. \citet{Jankovic2019} showed that, in an inner disc in which the gas accretion and the grain turbulent velocities are driven by the MRI, the grains can become small enough to escape the pressure trap through advection and radial mixing by the turbulent gas. Additionally, it was found that this leads to an enhanced dust-to-gas ratio of small dust grains throughout the inner disc, interior to the pressure maximum.

\citet{Jankovic2019} did not explicitly take into account the effects of dust on the MRI, whereas it can be expected that the accumulation of dust could quench the MRI in the innermost disc since dust grains adsorb free charges from the gas phase \citep{Sano2000, Ilgner2006, Wardle2007, Salmeron2008, Bai2009, Mohanty2013}. As a consequence of quenching the MRI, the strength of the turbulence would fall allowing some grain growth. However, this would concurrently push the MRI-active region and the pressure maximum inwards, possibly eliminating it from the inner disc. Evidently, the outcome is a function of the size and the abundance of the dust grains.

In a previous paper \citep[][Paper I]{Jankovic2021} we presented a model of a steady-state viscously accreting disc which includes both the MRI-driven viscosity and the effects of dust on the MRI. This accounts for the adsorption of free charges onto dust grains, and also for the electron (thermionic) and ion emission from dust grains into the gas phase. The thermionic and ion emission become important at temperatures above $\sim$1000\,K (so at the temperatures present in the inner disc) and act to increase the ionization fraction of the gas \citep{Desch2015}. 
We found that for 1\,$\mu$m grains, comprising 1\% of the disc mass, these dust effects balance out and result in a pressure maximum at roughly the same location as predicted from thermal ionization. Additionally, this model also self-consistently considers the disc opacity due to dust grains, thus accounting for the effects of dust on the disc thermal structure.

In the above work, we focused on a fiducial disc model. Building on this, in this paper we investigate how the inner disc structure changes with dust-to-gas ratio, dust grain size, and other disc and stellar parameters, in order to narrow down the region of parameter space where the formation of planetary cores in the inner disc is more likely. In section \ref{sec:theory} we discuss the theoretical expectations about the location of the pressure maximum based on the results of Paper I. We briefly overview our disc model in section \ref{sec:methods} and present our results in section \ref{sec:results}. In section \ref{sec:location_pressure_maximum} we focus on the existence and location of the gas pressure maximum as a function of the above parameters, exploring the entire parameter space in detail. In section \ref{sec:discussion} we discuss the implications of our results for the formation of the super-Earths and the limitations of our work, and in section \ref{sec:summary} we summarize our conclusions.

\section{Theoretical expectations} \label{sec:theory}
A local gas pressure maximum is expected to form in an accretion disc at steady-state if the material in the inner regions accretes faster than the material in the outer regions. In the context of viscous accretion discs, for the pressure maximum to form, the disc viscosity should decrease with distance from the star. If the viscosity is driven by the MRI, the requirement is for the coupling between the magnetic field and the gas (i.e., the ionization fraction) to decrease radially outwards.

Such a configuration has been predicted to arise in protoplanetary discs as the innermost regions can be hot enough to thermally ionize potassium, whereas outer regions are not \citep{Gammie1996}. \citet{Desch2015} have shown that, in fact, at the temperatures present in the inner disc, the dominant sources of ionization are thermionic and ion emission from small dust grains, and not thermal ionization. Nevertheless, the resulting ionization fraction sharply increases above a certain critical temperature, the same as in the case of thermal ionization. Moreover, for the materials out of which we expect the small grains to be made of in the inner disc, this critical temperature is likely  close to the temperature required for thermal ionization of potassium \citep[about 1000\,K; see the discussion in][]{Desch2015}. This suggests that the pressure maximum should arise at roughly the same distance from the star as in the case of thermal ionization. In Paper I we showed that, for fiducial stellar, disc and dust parameters, this is indeed the case. 

Therefore, we expect that a local gas pressure maximum indeed forms at the distance from the star at which the disc temperature reaches about $\sim$\,1000\,K. In the inner disc, the MRI is active around the disc midplane \citep[if the vertical temperature distribution is calculated self-consistently, see Paper I, and also][]{Terquem2008}, and so it is the disc midplane temperature that sets the location of the pressure maximum. Furthermore, as we show in Paper I, the midplane temperature is set by the heat released by the accretion and the disc cooling, while the heating by stellar irradiation has a small effect on the location of the pressure maximum, due to the inner disc being vertically optically thick. We also find that the inner disc is unstable to vertical convection.

Given the above findings, it is useful to consider how the disc midplane temperature (and thus the location of the pressure maximum) scales with different disc, stellar and dust parameters using the simple \citet{Shakura1973} viscous thin disc model, modified to account for vertical convection. In this model, the viscosity $\nu$ is parametrized by the dimensionless $\alpha$ parameter and given by $\nu=\alpha c_s^2 / \Omega$, where $c_s$ is the sound speed and $\Omega$ the angular Keplerian velocity. It is assumed that the disc is at steady-state, i.e., has a radially-constant gas accretion rate $\dot{M}$, and that the only source of heating is the accretion, that the disc is strongly optically thick (such that the majority of the discs mass is below the photosphere) and is convective up to the cooling radiation photosphere. Under these assumptions, \citet{Garaud2007} showed that the disc's midplane temperature ($T$) is approximately related to the effective temperature ($T_{\rm eff}$, determined by viscous dissipation) via $T{\rm \,}\propto{\rm \,}\tau_0^{2/7} T_{\rm eff}^{(7+2\beta)/7}$ where $\beta$ encapsulates the temperature dependence of the opacity via $\kappa = \kappa_0 (T/T_0)^\beta$, with $T_0$ a reference temperature and $\tau_0\equiv\kappa_0 \Sigma / 2$ (with $\Sigma$ the disc surface density) the optical depth to radiation at the reference temperature\footnote{The optical depth to cooling radiation can be calculated through the evaluation of $\tau=\int_0^\infty \kappa(T(z))\rho(z){\rm d}z$.}. For constant $\beta$ the disc structure is then described by a set of power-law relations. For the disc midplane temperature, away from the disc inner edge, we have
\begin{equation} \label{eq:ss_temp}
    T{\rm \,}\propto{\rm \,}\kappa_0^{2/9} \alpha^{-2/9} \dot{M}^{\frac{15+2\beta}{36}} M_*^{\frac{11+2\beta}{36}}  r^{-\frac{11+2\beta}{12}},
\end{equation}
where  $M_*$ the stellar mass, $\dot{M}$ the disc accretion rate and $r$ the cylindrical radius. Then, if the pressure maximum arises at a fixed temperature, its radial location scales with other parameters as:
\begin{equation} 
    r_{P_{\rm max}}{\rm \,}\propto{\rm \,}\kappa_0^{\frac{8}{3(11+2\beta)}} \alpha^{-\frac{8}{3(11+2\beta)}} \dot{M}^{\frac{15+2\beta}{3(11+2\beta)}} M_*^{1/3}.
\end{equation}
Inspection of our Rosseland mean opacities in Figure~\ref{fig:opacities} indicates values of $\beta\sim 0.5$ for temperatures around $1000~$K, thus for this choice of $\beta$ we find:
\begin{equation} \label{eq:ss_pmax}
    r_{P_{\rm max}}{\rm \,}\propto{\rm \,}\kappa_0^{2/9} \alpha^{-2/9} \dot{M}^{4/9} M_*^{1/3}.
\end{equation}
Therefore, for this choice of $\beta$, we obtain an expression that is exactly the same as the one found by \citet{Chatterjee2014}, who assumed that the disc was cooled radiatively. This agreement arises because for values of $\beta$ of order unity or smaller, the radiative $T(\tau)$ relation in discs $T\propto \tau^{1/4}T_{\rm eff}$ is a reasonable approximation for the mid-plane temperature in a very optically thick convective disc \citep[e.g.][]{Cannizzo1984}. Putting in the correct constants and fiducial disc parameters for a Solar mass star into eq. (\ref{eq:ss_pmax}), one arrives at a value of a few tenths of an AU, which falls within the range of the observed orbital radii of the close-in super-Earths \citep[see][]{Chatterjee2014}.

Equation (\ref{eq:ss_pmax}) tells us how the radius of the pressure maximum depends on the various disc and stellar parameters. If the pressure maximum arises at a fixed temperature, a disc that is hotter will feature a pressure maximum at a larger distance from the star, and a disc that is too cold will not feature a pressure maximum at all. The disc midplane is hotter if the heating rate is higher, or if radiative cooling is less efficient. The heating rate due to viscous dissipation is directly proportional to the accretion rate and the stellar mass. On the other hand, the cooling rate is reduced by increasing the optical depth of the disc. At a fixed value of viscosity parameter $\alpha$, higher accretion rate and stellar mass yield a disc with a higher surface density and thus a disc that is more optically thick and less efficient at cooling. Higher opacity also makes the disc more optically thick. Therefore, higher accretion rate, stellar mass and opacity all lead to a larger radius of the pressure maximum.

Conversely, higher value of the viscosity parameter $\alpha$ makes the disc surface density lower and thus it makes the disc less optically thick. The radius of the pressure maximum is thus inversely related to $\alpha$. Note that the value of $\alpha$ discussed here is the value at the location of the pressure maximum, i.e., at the location where the MRI is (largely) suppressed. It thus refers to the small viscosity driven either by propagation of turbulence from the adjacent MRI-active inner region or by other, hydrodynamical processes. In this work, it is a free parameter, and we refer to it as the minimum or dead-zone viscosity parameter.

Furthermore, we can expand on eq. (\ref{eq:ss_pmax}) by considering how the mass opacity ($\kappa$) depends on the properties of dust. First, we assume that the dust grain size distribution follows a power-law distribution $n(a)$\,$\propto$\,$a^{-3.5} da$ with a minimum grain size $a_{\rm min}$ and a maximum grain size $a_{\rm max}$. Second, we neglect gas opacities and scattering (or assume a constant albedo). Under these assumptions, the opacity can be approximated as the ratio of the surface area of all dust grains larger than the wavelength of peak local radiation (smaller grains, if there are any, contribute much less to the absorption cross section) and the total dust mass. For the above grain size distribution, this surface area is dominated by the grains of the size of the peak wavelength, while the mass is dominated by the largest grains. Then, if we also assume that the maximum dust grain size ($a_{\rm max}$) is much larger than the peak wavelength, at a fixed temperature, the opacity $\kappa$ has a simple dependence on $a_{\rm max}$ and the dust-to-gas ratio $f_{\rm dg}$, $\kappa_0{\rm \,}\propto{\rm \,}f_{\rm dg} a_{\rm max}^{-1/2}$. Finally, for the radial location of the pressure maximum we obtain
\begin{equation} \label{eq:ss_pmax_dust}
    r_{P_{\rm max}}{\rm \,}\propto{\rm \,}f_{\rm dg}^{2/9} a_{\rm max}^{-1/9} \alpha^{-2/9} \dot{M}^{4/9} M_*^{1/3}.
\end{equation}

Evidently, this scaling does not take into account how the critical ionization temperature (at which thermionic and ion emission become efficient) depends on the dust properties. Even if the disc is thermally ionized, the above derivation neglects the dependence of the critical temperature on the density and the complexity of the criteria for the onset of the MRI. However, for the case of thermal ionization, we can compare eq. (\ref{eq:ss_pmax_dust}) with fits to the numerical results of \citet{Mohanty2018}, who self-consistently coupled the simple Shakura-Sunyaev disc model with thermal ionization of potassium and a detailed prescription for MRI-driven viscosity (the same one used in this work). The fits were obtained for the viscosity parameter, accretion rate and stellar mass, which combined together yield
\begin{equation} \label{eq:moh_rpmax}
    r_{P_{\rm max}}{\rm \,}\propto{\rm \,}\alpha^{-1/4} \dot{M}^{1/2} M_*^{1/3}.
\end{equation}
While the exponents here deviate somewhat from eq. (\ref{eq:ss_pmax_dust}), the deviations are small \citep[see also][]{Kretke2009}. Therefore, for the case of thermal ionisation, the scaling of $r_{P_{\rm max}}$ with $\alpha$, $\dot{M}$ and $M_*$ given in eq. (\ref{eq:ss_pmax_dust}) remains approximately correct. In this work, we use our new model to determine whether these deviations from the simple scaling become larger when thermionic and ion emission are accounted for, and also explore the dependence on the dust parameters.

Finally, note that the entire discussion so far has been made under an assumption that the disc's ionization fraction rises sharply above a fixed critical temperature. However, non-thermal sources of ionization are also present in protoplanetary discs \citep[most importantly stellar X-rays;][]{Glassgold1997,Ercolano2013}. Non-thermal sources of ionization increase the ionization fraction at all orbital radii, not only in the high-temperature inner region. In Paper I we explored how this affects the MRI-driven viscosity in the inner disc, for a set of fiducial model parameters. The effect of any of the ionisation sources on the MRI-driven viscosity depends on the magnetic field strength. Under the assumptions made in Paper I about the field strength (that it is vertically constant and evolves to maximize the MRI-driven $\alpha$), the stellar X-rays only contribute to the MRI-driven accretion in the outer regions, outwards from the pressure maximum, and also at the location of the pressure maximum, since $\alpha$ has to be a continuous function. If this is indeed the case, the role of the X-rays is simply to increase the $\alpha$ in eq. (\ref{eq:ss_pmax_dust}), which pushes the pressure maximum inwards. Since the contribution of the X-rays to accretion in the inner regions is neglected, the above assumptions maximize the effect of the X-rays on the location of the pressure maximum. Nevertheless, in Paper I we showed that, for fiducial disc and stellar parameters, this effect is completely negligible. Firstly, for those fiducial parameters, the surface densities in the inner disc are much higher than the column that the X-rays can penetrate. Second, in the X-ray ionized regions, dust grains act as an efficient recombination pathway for the ions, which significantly suppresses the MRI \citep{Sano2000, Ilgner2006, Wardle2007, Salmeron2008, Bai2009, Mohanty2013}. The adsorption of gas-phase ions onto the dust grain surfaces depends primarily on the total dust grain surface area. Therefore, we can expect that the stellar X-rays become more important in a less massive disc (e.g., for disc accretion rates that are lower than our fiducial one), and also for lower dust-to-gas ratios and larger dust grains. As we explore the disc parameter space in this paper, we identify the limits in which the ionization due to stellar X-rays becomes important for the location (and the existence) of the pressure maximum.

\section{Methods} \label{sec:methods}
Our disc model is presented in Paper I. Here, we only summarize the main points. It is assumed that the viscously-accreting disc is in steady-state, i.e., that the gas accretion rate $\dot{M}$ is radially constant. The disc structure is calculated self-consistently with disc opacities, ionization state and the viscosity due to the MRI. The disc is assumed to be in vertical hydrostatic and thermal equilibrium, heated by viscous dissipation and stellar irradiation, and cooled radiatively and/or via convection.

We account for heating by stellar irradiation using the grazing angle prescription \citep{Calvet1992,Chiang1997,DAlessio1999}. For the first radial point we assume the flat-disc approximation. The disc structure and the grazing angle at every other radial point are determined self-consistently. To calculate the grazing angle, we perform a smoothing over a number of grid points (see Paper I for details). As such, the first few grid points in our calculation are sensitive to how we perform this smoothing, and we indicate the affected region in our figures throughout this paper.

\begin{figure}
    \centering
    \includegraphics[width=\columnwidth, trim={0 3cm 0 4cm}, clip]{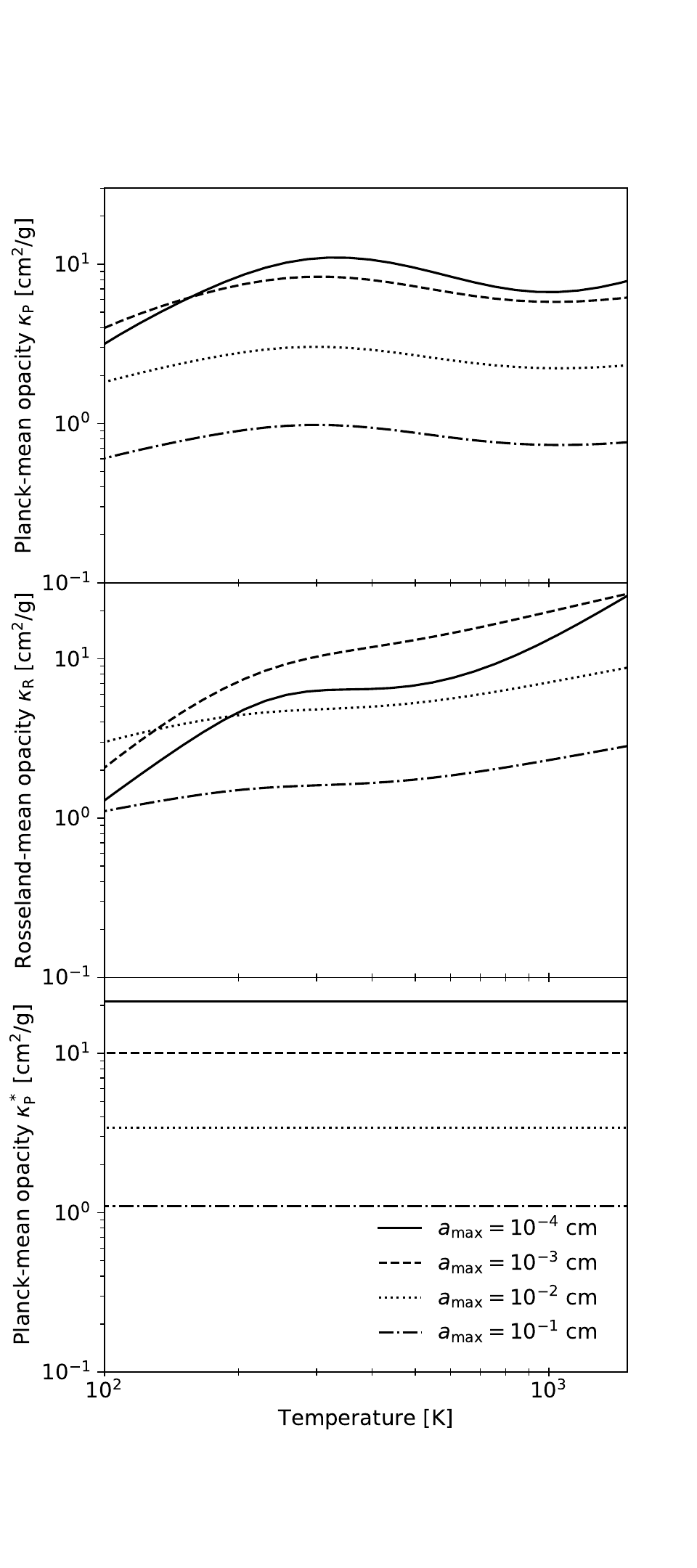}
    \caption{Planck-mean opacity $\kappa_{\rm P}$ (top), Rosseland-mean opacity $\kappa_{\rm R}$ (middle), and Planck-mean opacity at the stellar effective temperature $\kappa_{\rm P}^*$ (i.e., the absorption coefficient for the stellar irradiation; bottom), per unit mass of gas, as functions of disc temperature, for different maximum grain sizes $a_{\rm max}$ as indicated in plot legend, assuming a dust-to-gas ratio $f_{\rm dg}=0.01$ and stellar effective temperature $T_*=4400$\,K. Absorption is dominated by small grains, and so the Planck-mean opacities decrease with increasing maximum grain size. The Rosseland-mean opacity is a non-monotonic function of grain size for $a_{\rm max} \lesssim 10^{-2}$\,cm. 
    }
    \label{fig:opacities}
\end{figure}

We consider disc opacities due to silicate dust grains, and we calculate the opacities for a MRN grain size distribution \citep{Mathis1977}, with a minimum grain size $a_{\rm min}=0.1$\,$\mu$m and a maximum grain size $a_{\rm max}$, using optical constants from \citet[][see Paper I for details]{Draine2003}, and assuming bulk density of dust grains $\rho_{\rm s}=3.3$\,g\,cm$^{-3}$. 
In the calculation of the Rosseland-mean opacity we include the scattering coefficient, but assume that the scattering is isotropic. Figure \ref{fig:opacities} shows the opacities, calculated per unit mass of gas, for different maximum grain sizes $a_{\rm max}$, assuming a dust-to-gas ratio $f_{\rm dg}=0.01$ and stellar effective temperature $T_*=4400$\,K. 
We only consider the structure of the disc beyond the silicate sublimation line, when dust opacities typically dominate over those of the gas, and so neglect the opacities due to gas molecular and atomic lines (but see section \ref{sec:limitations}). Additionally, the water ice and carbonaceous grains have sublimation temperatures that are much lower than the temperatures expected in the hot MRI-active regions \citep[e.g.][]{Pollack1994}. These grains may condense in the colder layers such as the photosphere to the disc radiation. Still, for simplicity, we neglect the contribution from other dust species.

The disc ionization state is calculated using a simple chemical network \citep{Desch2015} that includes thermal (collisional) ionization of potassium; ionization of molecular hydrogen by stellar X-rays, cosmic rays and radionuclides (producing metal (magnesium) ions by charge transfer); gas-phase recombinations; adsorption onto dust grains and thermionic and ion emission from dust grains. Both charged and neutral species can be adsorbed (or condensed) in collisions with dust grains. Neutral potassium atoms, potassium ions and electrons can also be emitted back into the gas phase. The rates at which the latter processes occur increase with increasing temperature, as determined by 3 different activation energies and the charge state of the dust grains. Firstly, evaporation of potassium atoms from dust grains is determined by our adopted binding energy of potassium ($E_{\rm a}=3.26$\,eV), whose value is chosen to match the condensation temperature of a common potassium-bearing mineral \citep[see][]{Desch2015}. Furthermore, a fraction of the potassium atoms evaporating from dust grains may be ionised in the process. This is referred to as ion emission, and it is a function of the ionization potential ($\textrm{IP}=4.34$\,eV) of potassium, as well as the work function of the material of which the dust is composed \citep[for the discussion of the adopted value, $W=5$\,eV, see][]{Desch2015}. Thermionic emission, i.e., emission of electrons from heated dust grains, is also determined by the work function. Additionally, the charge state of the grains effectively changes the work function. For example, ion emission results in negatively charged grains, which reduces the effective work function and increases the thermionic emission rate.

Only a single dust grain species of size $a_{\rm gr}=0.1$\,$\mu$m is considered in the chemical network, but an effective dust-to-gas ratio $f_{\rm eff}$ is chosen to mimic the full size distribution stated above (see Paper I for details). 
This approach is based on an assumption that multiple grain species behave independently in the chemical network, and that it is a combination of the grain abundance and the grain size that controls the chemistry of the disc \citep[see also][]{Bai2009}. Specifically, we choose $f_{\rm eff}$ here such that the ionisation threshold temperature (above which thermionic and ion emission become efficient) approximately mimics that expected for the full range of grain sizes (see Paper I). 

Among the non-thermal sources of ionization, stellar X-rays appear to be the most important one for the models investigated here where the focus is on the inner regions of the disc. The ionization rate of molecular hydrogen by stellar X-rays is calculated using \citet{Bai2009} fits to the \citet{Igea1999} Monte Carlo simulations (using the fits for $k T_X = 3$\,keV), so at any point in the disc it is a function of the cylindrical radius and the vertical mass column from the disc surface to that point. We ignore the X-rays coming through the bottom side of the disc and note that in a low-surface-density disc this may increase the ionization fraction by a factor of 2 at most. For the stellar X-ray luminosity we adopt $L_{\rm X} = 10^{-3.5} L_{\rm bol}$ \citep[e.g.][but note that there is significant scatter in the observed luminosities and for a single star this luminosity may also be variable due to stellar flares]{Wright2011, Gudel2007, Preibisch2005}.

Finally, the viscosity due to the MRI, parametrized using the \citet{Shakura1973} $\alpha$ parameter, is calculated using a prescription based on the results of magnetohydrodynamic simulations (see Paper I). This accounts for the suppression of the MRI by Ohmic and ambipolar diffusion. In the MRI-dead zones (where the MRI is suppressed), we assume the gas can still accrete due to a small constant viscosity parameter $\alpha_{\rm DZ}$, induced either by the adjacent MRI-active zone or by purely hydrodynamical instabilities. The viscous $\alpha$ is calculated both as a function of radius and height, and we define a vertically-averaged viscosity parameter
\begin{equation} \label{eq:alphabar}
    \bar\alpha \equiv \frac{ \int_0^{z_{\rm surf}} \alpha P dz }{ \int_0^{z_{\rm surf}} P dz },
\end{equation}
where $z_{\rm surf}$ is the height of the disc surface, defined as the height above the mid-plane where the gas pressure falls below a small constant value ($P(z_{\rm surf}) = 10^{-10} \textrm{\,dyn\,cm}^{-2}$). The weighting of the viscosity parameter $\alpha$ by pressure is motivated by the relationship between the $\alpha$ and the accretion rate: in steady-state, the accretion rate is proportional to the numerator in the above expression. 

\section{Results} \label{sec:results}
Here, we explore the effects of varying the important physical parameters on the inner disc structure. Initially, we vary the parameters one-by-one: dust-to-gas ratio $f_{\rm dg}$ and maximum dust grain size $a_{\rm max}$ in section \ref{sec:ratio_size}, and the gas accretion rate $\dot{M}$, stellar mass $M_*$, and the dead-zone viscosity $\alpha_{\rm DZ}$ in section \ref{sec:mdot_mstar_alphadz}. We then extend the exploration of the parameter space by concurrently considering a larger dust grain size and a lower gas accretion rate, in section \ref{sec:big_dust_low_mdot}, discovering an X-ray dominated solution. As our fiducial model, taken from Paper I, we consider a disc with a gas accretion rate $\dot{M}=10^{-8}$\,M$_\odot$\,yr$^{-1}$, stellar mass $M_*=1$\,M$_\odot$, stellar radius $R_*=3$\,R$_\odot$, effective stellar temperature $T_*=4400$\,K\footnote{These specific stellar parameters correspond to a Solar-mass star at an age of 0.5\,Myr in the stellar evolution models of \citet{Baraffe2015}. Over the first 5\,Myr the luminosity of a Solar-mass star decreases by a factor of 5 in these models, and so the adopted parameters are roughly valid throughout the disc lifetime. 
}, viscosity parameter in the MRI-dead zone $\alpha_{\rm DZ}=10^{-4}$, dust-to-gas ratio $f_{\rm dg}=10^{-2}$, and maximum dust grain size $a_{\rm max}=10^{-4}$\,cm. In Paper I, this model was used to discuss the impact of various physical and chemical processes on the inner disc structure.

\subsection{Dust-to-gas ratio and dust size} \label{sec:ratio_size}
As noted in section \ref{sec:theory}, dust has two effects on the disc structure in our model: it determines opacities in the disc, and it affects the disc ionization state. 
To better understand the results of varying dust properties, we first consider only the dust opacities. That is, in section \ref{sec:therm_ratio_size}, we consider a model with a vastly simplified chemical network, in which the only source of ionization is thermal ionization of potassium and free charges recombine only in the gas phase. Then, in section \ref{sec:chem_ratio_size} we present the results of our full model which also accounts for the adsorption of charges onto dust grains, thermionic and ion emission from dust grains, and ionization of molecular hydrogen by stellar X-rays, cosmic rays and radionuclides.

\subsubsection{Thermally-ionized model} \label{sec:therm_ratio_size}
\begin{figure*}
    \centering
    \includegraphics[width=0.47\textwidth,trim={0 4.5cm 0 4.5cm},clip]{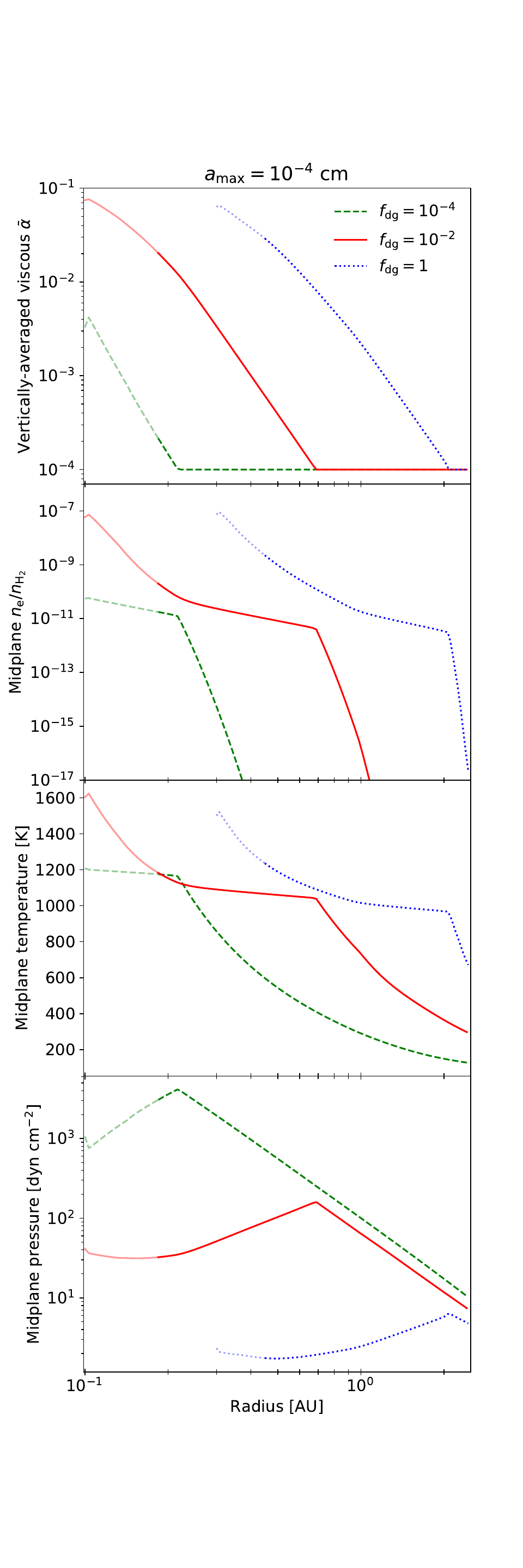}
	\includegraphics[width=0.47\textwidth,trim={0 4.5cm 0 4.5cm},clip]{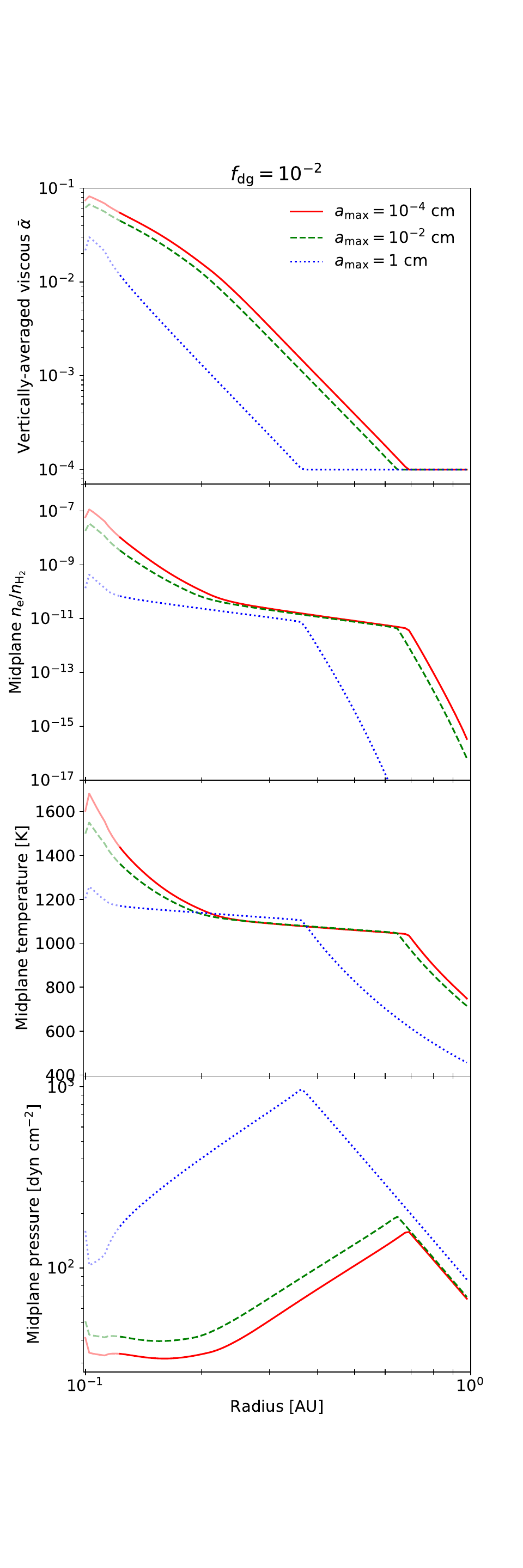}
    \caption{Results for a thermally-ionized disc (where the only source of ionization is thermal ionization of potassium, and dust effects on the disc chemistry are not included). The left column shows models with a constant maximum grain size $a_{\rm max}=10^{-4}$\,cm and varying dust-to-gas ratio $f_{\rm dg}$ as indicated in plot legend. The right column shows models with a constant dust-to-gas ratio $f_{\rm dg}=10^{-2}$ and varying maximum grain size $a_{\rm max}$ as indicated in plot legend. The rows show radial profiles of (from top to bottom) vertically-averaged viscosity parameter $\bar\alpha$, midplane free electron fraction $n_{\rm e}/n_{\rm H_2}$, midplane temperature and midplane pressure. The inner edge of the $f_{\rm dg}=1$ model is set to $\sim 0.3$\,AU, since radially inwards temperature increases above the sublimation temperature of silicates. The light lines indicate the regions affected by the inner boundary condition (see Section \ref{sec:methods}). The radius of the pressure maximum is larger for a larger dust-to-gas ratio and a smaller maximum grain size. Note that the axis ranges are different in the left and the right column. See Section \ref{sec:therm_ratio_size}.}
    \label{fig:thermal_ratio_size}
\end{figure*}

In this section we consider a model which includes dust opacities, but does not include dust effects on the disc chemistry, nor ionization of molecular hydrogen. Thus, the ionization fraction is set exclusively by thermal ionization. The results of varying the dust-to-gas ratio, $f_{\rm dg}$, in the range $10^{-4}-1$ are shown in the left column of Fig. \ref{fig:thermal_ratio_size}.

In this simplified, thermally-ionized disc model, the MRI is only active at small radii. Therefore, the viscosity is highest in the innermost region where the midplane ionization fraction and the midplane temperature (shown in the second and third row, respectively) are highest, and it decreases with distance from the star. At some radius the ionization fraction drops below that needed to sustain the MRI, and the viscosity parameter $\bar\alpha$ falls to the minimum, dead-zone value $\alpha_{\rm DZ}$. That radius is the location of the local gas pressure maximum, shown in the bottom panel.

We find that a higher dust-to-gas ratio results in a larger MRI-active zone (the region where $\bar\alpha>\alpha_{\rm DZ}$). This is the behaviour expected from eq. (\ref{eq:ss_pmax_dust}). Because the inner disc is optically-thick, the disc midplane temperature is set by the accretion heat released near midplane and the optical depth of the disc to its own radiation (with a caveat that the vertical temperature gradient is additionally limited by convection; see Paper I). The disc's opacity is directly proportional to the dust-to-gas ratio, i.e., a disc with more dust is more optically-thick. 
As discussed in section 2, increasing optical depth makes the cooling less efficient, and the midplane hotter and more ionized, leading to a higher MRI-induced viscosity at a given radius. At lower dust-to-gas ratios gas opacities (which are neglected here) should become more important than at higher dust-to-gas ratios, but the main results shown here would likely be unaffected if the gas opacities were included (see Section \ref{sec:limitations}).

Furthermore, the right column of Fig. \ref{fig:thermal_ratio_size} shows models with a constant dust-to-gas ratio of $f_{\rm dg}=10^{-2}$, but varying the maximum dust grain size $a_{\rm max}$ in the range $10^{-4}-1$\,cm. For the three values of $a_{\rm max}$ considered here, the MRI-active region becomes smaller if dust grains are larger. This is because, for these values of $a_{\rm max}$, larger dust grains have lower opacities (see Fig. \ref{fig:opacities}), making the inner disc less optically thick. Just as in a disc with a lower dust-to-gas ratio, this makes the disc midplane cooler and less ionized. 

Therefore, in these simplified, thermally-ionized models, if dust growth were to happen, this would result in the MRI-active zone edge being pushed inwards. We emphasize that it is the changes to the dust opacities alone that lead to these significant changes in the extent of the MRI-active zone. The effects of including dust grains in our chemical network are explored next.

\subsubsection{Full model} \label{sec:chem_ratio_size}
\begin{figure*}
    \centering
    \includegraphics[width=0.47\textwidth,trim={0 4.5cm 0 4.5cm},clip]{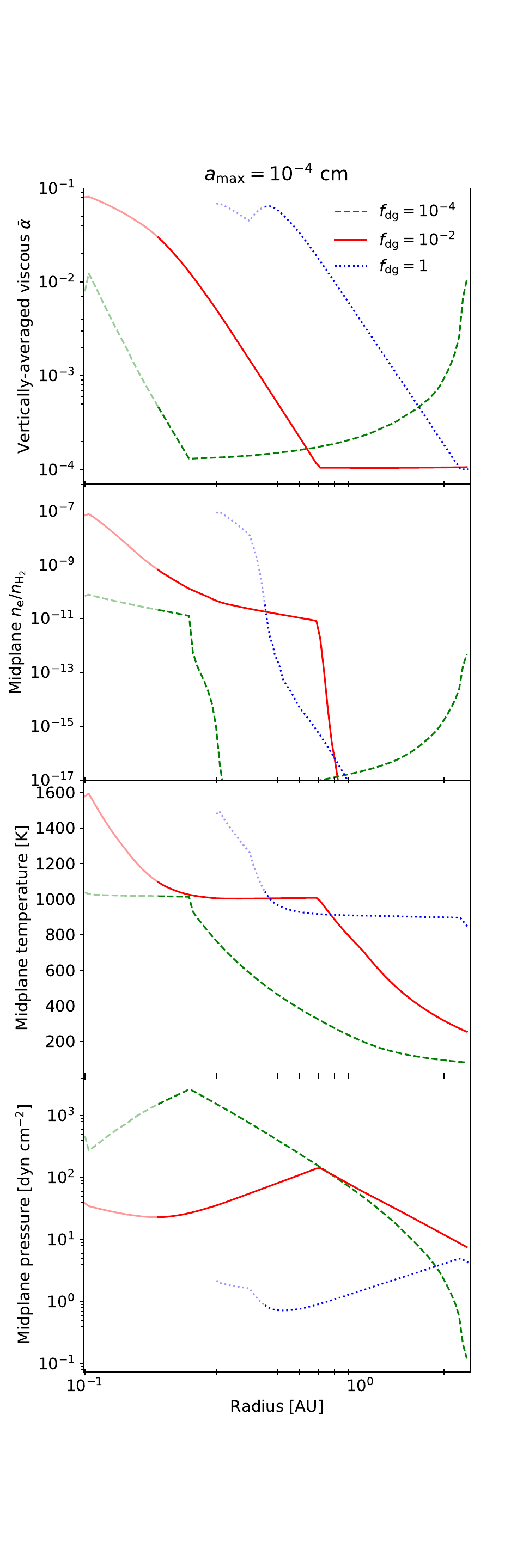}
	\includegraphics[width=0.47\textwidth,trim={0 4.5cm 0 4.5cm},clip]{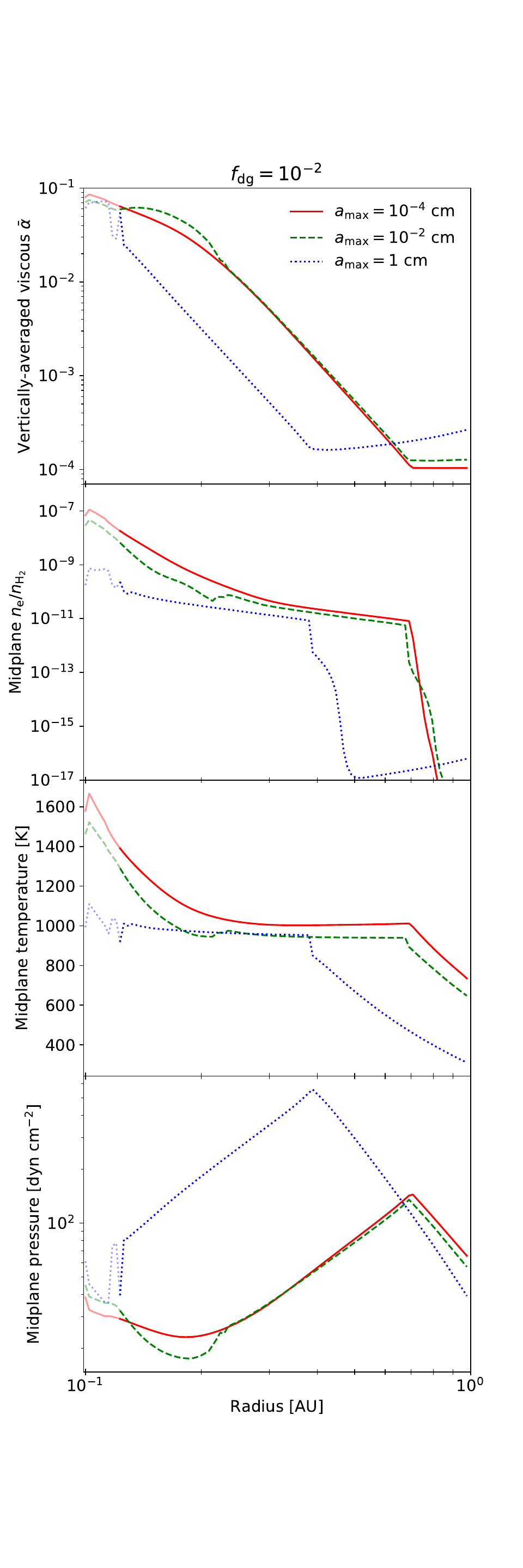}
    \caption{Results for our full model which includes dust effects on disc chemistry and ionization of molecular hydrogen. The left column shows models with a constant maximum grain size $a_{\rm max}=10^{-4}$\,cm and varying dust-to-gas ratio $f_{\rm dg}$ as indicated in plot legend. The right column shows models with a constant dust-to-gas ratio $f_{\rm dg}=10^{-2}$ and varying maximum grain size $a_{\rm max}$ as indicated in plot legend. The rows show radial profiles of (from top to bottom) vertically-averaged viscosity parameter $\bar\alpha$, midplane free electron fraction $n_{\rm e}/n_{\rm H_2}$, midplane temperature and midplane pressure. The light lines indicate the regions affected by the inner boundary condition (see Section \ref{sec:methods}). The radius of the pressure maximum is larger for a larger dust-to-gas ratio; it is approximately the same for the maximum grain size of $10^{-4}$\,cm and $10^{-2}$\,cm, but much smaller for the maximum grain size of $1$\,cm. Note that the axis ranges are different in the left and the right column. See Section \ref{sec:chem_ratio_size}.}
    \label{fig:chem_ratio_size}
\end{figure*}

In this section, we consider our full model that additionally includes (direct) effects of dust on the ionization fraction (adsorption of free charges onto dust grains, thermionic and ion emission), and also ionization of molecular hydrogen. Three sources of ionization are considered for the latter (stellar X-rays, cosmic rays and radionuclides), the X-rays being the most important (see Paper I). The results of varying the dust-to-gas ratio and maximum dust grain size in this full model are shown in the left and the right column of Fig. \ref{fig:chem_ratio_size}, respectively. 

As in the simplified, thermally-ionized models discussed above, in the innermost regions the viscosity parameter decreases with distance from the star. Here, however, the viscosity parameter reaches a minimum value close to the dead-zone viscosity, and then increases again radially outwards, due to ionization by stellar X-rays (this is true in all models, even if not always evident in the plots). As discussed in Section~2 \citep[Paper I;][]{Desch2015}, in the full model the main source of ionization are thermionic and ion emission from dust grains in the inner disc. While this is a fundamentally different mechanism from gas-phase thermal ionization, the ionization states of the disc are quantitatively similar in the two cases due to their similar activation energies. As a result, the viscosity parameter in these innermost regions is similar to the models with no dust in the chemical network. In Paper I, we discussed this case for the fiducial maximum grain size $a_{\rm max}=10^{-4}$\,cm. The results presented in this work show that this conclusion holds at a wide range of dust-to-gas ratios and grain sizes (e.g., compare Fig. \ref{fig:chem_ratio_size} with Fig. \ref{fig:thermal_ratio_size}). 

\begin{figure}
    \centering
    \includegraphics[width=0.49\textwidth]{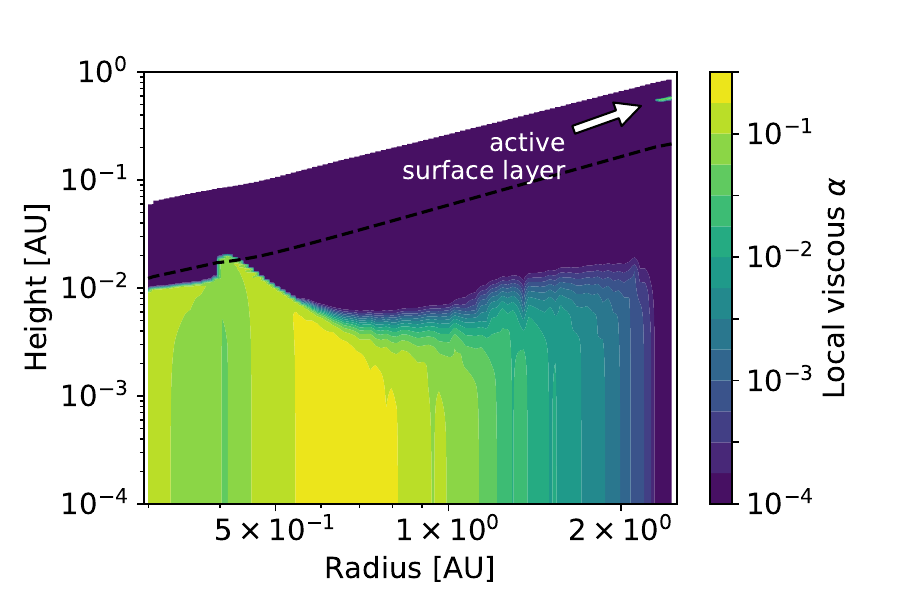}
    \includegraphics[width=0.49\textwidth]{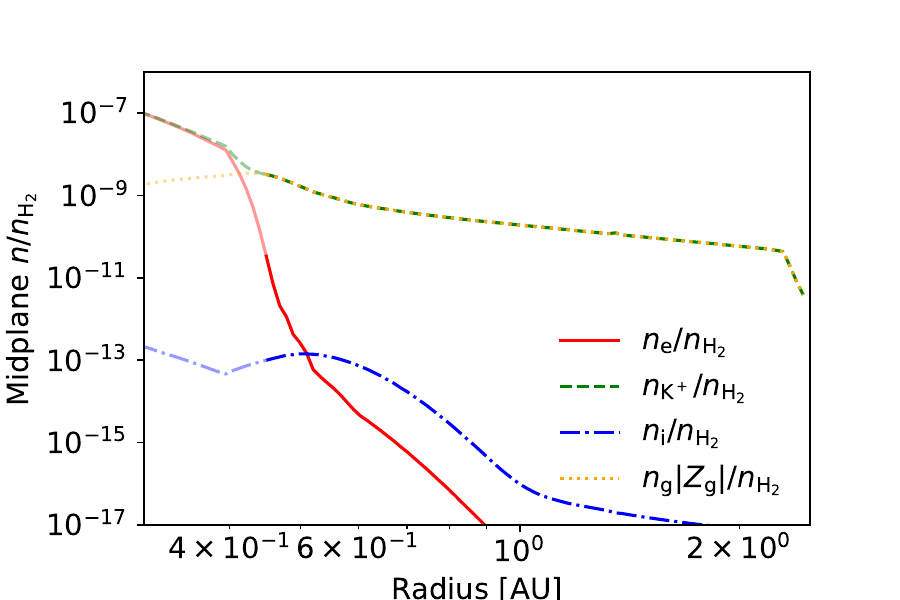}
    \caption{Local viscosity parameter $\alpha$ as a function of location in the disc (top) and the fractional number densities of various ionized species at the disc midplane as a function of radius (bottom), in a model with a dust-to-gas ratio of $f_{\rm dg}=1$ (with other parameters equal to the fiducial values). The species shown in the bottom panel are free electrons ($n_{\rm e}$), potassium ions ($n_{\rm K^+}$), magnesium ions ($n_{\rm i}$) and dust grains ($n_{\rm g}$, carrying a mean charge $Z_{\rm g}$). See Section \ref{sec:chem_ratio_size}.}
    \label{fig:model_13}
\end{figure}

The case of high dust-to-gas ratio of $f_{\rm dg}=1$ deviates somewhat from the above scenario. In this model, the midplane free electron fraction decreases substantially already at the distance of $\sim 0.9$\,AU (see the left panel, second row in Fig. \ref{fig:chem_ratio_size}). However, the viscous $\bar\alpha$ remains high out to $\sim 2$\,AU in this model (the top left panel in Fig. \ref{fig:chem_ratio_size}). In this high-$\bar\alpha$ region, the MRI is indeed active at the disc midplane (as shown in the top panel of Fig. \ref{fig:model_13}), despite the large decrease in the midplane electron number density. What drives the MRI in this case? 

The bottom panel of Fig. \ref{fig:model_13} shows that between $\sim 0.4$\,AU and $\sim 2$\,AU the main ionized species are the potassium ions and the dust grains. As expected, the number density of electrons decreases with increasing dust-to-gas ratio (keeping other parameters fixed). However, the opposite is true for the number density of potassium ions evaporating from dust grains, above $\sim 900$\,K \citep[see][]{Desch2015}, which increases with increasing dust density. In the resulting disc ionization state, due to charge conservation, the total charge of potassium ions equals the total charge on dust grains. Clearly, since the dust grains have a much higher inertia than potassium ions, it is the potassium ions that couple the gas to the magnetic field. Overall, these results show that emission of potassium from dust grains is sufficient to sustain the MRI out to large radii, at high dust-to-gas ratios. Although, we note that at dust-to-gas ratios approaching unity the dynamical back reaction of the dust on the gas would need to be included, an effect which is poorly studied in the context of MRI turbulence and by extension not included in our parameterisation of the how the viscosity depends on the ionization structure. Clearly, MHD simulations of this disc state are warranted to study its behaviour. Particularly as our previous work in \citet{Jankovic2019} has indicated dust-to-gas ratios approaching unity in the inner MRI active regions is a possible outcome of disc evolution. 

Furthermore, as noted above, in our full model the viscosity parameter $\bar\alpha$ reaches a minimum value, outwards from which it increases with radius. The minimum in the viscosity parameter corresponds to the location of the gas pressure maximum, shown in the bottom panels of Fig. \ref{fig:chem_ratio_size}. Outwards of the pressure maximum, the temperature at the disc midplane is too low for efficient ionization, and so the MRI is only active in an X-ray-ionized layer high {\it above} disc midplane (seen near $\sim 2$\,AU in the top panel of Fig. \ref{fig:model_13}). Inwards of the pressure maximum, this X-ray-ionized layer does not appear; as discussed in Paper I, this is due to our assumption that the magnetic field strength is vertically constant, and the fact that the magnetic field strengths required for the MRI differ greatly in the high-density midplane and the low-density upper layers. In the outer, X-ray-ionized regions, the viscosity parameter increases with decreasing dust-to-gas ratio and increasing dust grain size. This is because the main source of ionization are the stellar X-rays, and dust only acts as a recombination pathway. For lower dust-to-gas ratios and, equivalently, higher maximum grain size, the total grain surface area (onto which free charges adsorb) decreases, leading to higher ionization fraction and higher MRI-driven viscosity \citep{Sano2000, Ilgner2006}. As expected from eq. (\ref{eq:ss_pmax_dust}), this increase in the value of $\bar\alpha$ at the location of the pressure maximum pushes the pressure maximum inwards. This effect is somewhat exaggerated since the effect of the X-rays on $\bar\alpha$ is neglected inwards of the pressure maximum (due to the model limitations noted above). Nevertheless, for the majority of the parameter space explored so far the contribution to the accretion rate from the X-ray-ionized layer is low in all cases and the viscosity parameter $\bar\alpha \sim \alpha_{\rm DZ}$ in the outer regions. In other words, in these outer regions the gas primarily accretes through the dead zone, and the stellar X-rays do not affect strongly the location of the pressure maximum (for the given gas accretion rate).

Overall, the extent of the high-viscosity inner region and the location of the pressure maximum is dictated by the dependence of disc's vertical optical thickness on the dust-to-gas ratio and dust grain size through the effects discussed in the previous section. In our full model the location of the pressure maximum is approximately the same for $a_{\rm max}=10^{-4}$\,cm and $a_{\rm max}=10^{-2}$\,cm (see the top right panel in Fig. \ref{fig:chem_ratio_size}). This is because a larger grain size results in a lower effective dust-to-gas ratio in our chemical network. This slightly decreases the critical temperature at which the thermionic and ion emission make the gas sufficiently ionized to start the MRI at disc midplane. Concurrently, the dust opacity for the case of $a_{\rm max}=10^{-2}$\,cm is only slightly smaller than for $a_{\rm max}=10^{-4}$\,cm in the relevant temperature range (see Fig. \ref{fig:opacities}). The location of the pressure maximum as a function of dust grain size is considered in more detail in section \ref{sec:location_pressure_maximum}.

\subsection{Gas accretion rate, stellar mass, dead-zone viscosity} \label{sec:mdot_mstar_alphadz}
\begin{figure}
    \centering
	\includegraphics[width=0.47\textwidth,trim={0 4.5cm 0 4.5cm},clip]{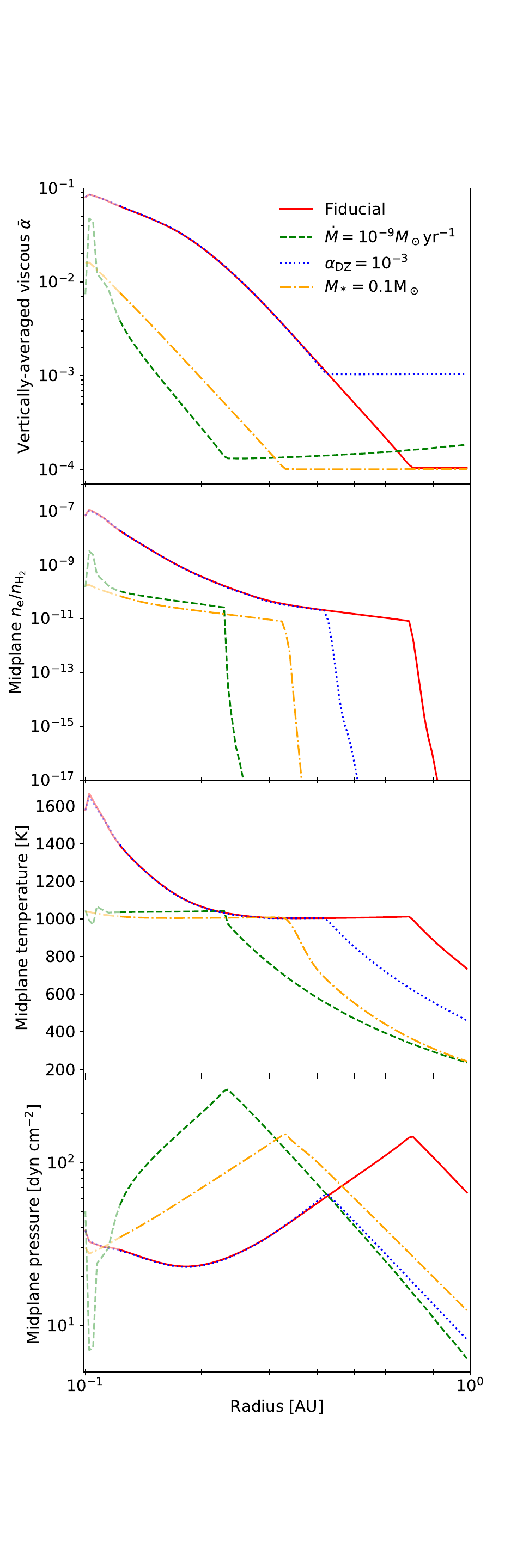}
    \caption{Results for our full model which includes dust effects on disc chemistry and ionization of molecular hydrogen. The different panels show radial profiles of (from top to bottom) vertically-averaged viscosity parameter $\bar\alpha$, midplane free electron fraction $n_{\rm e}/n_{\rm H_2}$, midplane temperature and midplane pressure. Fiducial model (solid lines) has a gas accretion rate $\dot{M}=10^{-8}$\,M$_\odot$\,yr$^{-1}$, stellar mass $M_*=1$\,M$_\odot$ and dead-zone viscosity $\alpha_{\rm DZ}=10^{-4}$. Models with a lower gas accretion rate (dashed line), a smaller stellar mass (dash-dotted line) or a larger dead-zone viscosity (dotted line) all yield the gas pressure maximum at a smaller radius. The light lines indicate the regions affected by the inner boundary condition (see Section \ref{sec:methods}). See Section \ref{sec:mdot_mstar_alphadz}.}
    \label{fig:mdot_mstar_alphadz}
\end{figure}

In this section we keep the dust-to-gas ratio and the maximum dust grain size constant and equal to our fiducial values, and investigate how the structure of the inner disc changes with varying gas accretion rate, stellar mass and dead-zone viscosity. Fig. \ref{fig:mdot_mstar_alphadz} shows the results of our fiducial model compared to three other models in which we vary these three parameters. The different panels show, from top to bottom, the vertically-averaged viscosity parameter ($\bar\alpha$), midplane free electron fraction ($n_{\rm e}/n_{\rm H_2}$), midplane temperature and midplane pressure, as functions of radius.

In each panel the dashed line shows a model with a gas accretion rate $\dot{M}=10^{-9}$\,M$_\odot$\,yr$^{-1}$, lower than in our fiducial model with $\dot{M}=10^{-8}$\,M$_\odot$\,yr$^{-1}$, shown by the solid line. 
The lower gas accretion rate results in a smaller high-viscosity inner region, and a gas pressure maximum at a shorter radius. This is in line with the theoretical expectations discussed in Section \ref{sec:theory}. In the optically-thick inner disc, the midplane temperature (and consequently the ionization fraction and the viscosity) is set by the rate of viscous dissipation, and the latter is directly proportional to the gas accretion rate. Additionally, fixing other parameters, gas accretion rate also sets the gas surface density, and thus the disc optical depth and how efficiently the disc cools. Therefore, the lower gas accretion rate yields a colder, less ionized disc. 
The radius of the gas pressure maximum scales with the gas accretion rate approximately as $r_{P{\rm max}} \propto \dot{M}^{1/2}$. This is close to the prediction given by eq. (\ref{eq:ss_pmax_dust}) and, in fact, the same as the scaling given by eq. (\ref{eq:moh_rpmax}), previously found by \citet[][]{Mohanty2018}, who neglected heating by stellar irradiation, dust effects and ionization of molecular hydrogen. While we find that stellar irradiation and ionization of molecular hydrogen are indeed unimportant for the model parameters chosen here, the dust effects are not. However, as discussed in Section \ref{sec:theory}, while the chemistry setting the ionization state of the disc is qualitatively different in their simple models (thermal ionization of potassium) and in the models presented here (thermionic and ion emission), in both cases the ionization fraction increases sharply above roughly the same temperature ($\sim 1000$\,K), yielding the same approximate scaling.

X-ray ionization of molecular hydrogen becomes more important at the lower gas accretion rate. X-rays activate the MRI in a layer high above the disc midplane outwards of the pressure maximum, increasing the viscosity parameter $\bar\alpha$ compared to the dead-zone value $\alpha_{\rm DZ}$. For an accretion rate of $10^{-9}$~M$_\odot$~yr$^{-1}$ the vertically averaged viscosity has increased by a factor of 2 outside the dead-zone over the model with an accretion rate of $10^{-8}$~M$_\odot$~yr$^{-1}$. This is because in these outer regions a lower gas accretion rate results in lower gas surface densities. Since the accretion rate carried by the X-ray ionized layer is roughly constant (e.g. set by the penetration depth of the X-rays) at lower accretion rates it has a larger relative contribution to the total accretion rate. This contribution remains small for the small maximum grain size assumed here; in the next section we discuss how this finding changes if grains are larger.

We find that the structure of the inner disc surrounding a $M_*=0.1$\,M$_\odot$ star\footnote{Here we adopt a stellar radius of $R_*=1$\,R$_\odot$ and an effective temperature of $T_*=2925$\,K. 
} (shown by the dash-dotted lines in Fig. \ref{fig:mdot_mstar_alphadz}) is merely shifted radially inwards compared to our fiducial model with $M_*=1$\,M$_\odot$. 
Similar to the gas accretion rate, the stellar mass determines the disc heating rate due to accretion, as well as the disc optical depth through the dependence of the disc surface density on the stellar mass. The resulting approximate scaling $r_{P_{\rm max}} \propto M_*^{-1/3}$ is identical to the one given by eq. (\ref{eq:ss_pmax_dust}), stressing again that (for the chosen $\dot{M}$, dust parameters etc.) stellar irradiation and ionization by stellar X-rays are unimportant in setting the location of the pressure maximum. Note that in Fig. \ref{fig:mdot_mstar_alphadz} we vary the disc parameters one-by-one from our fiducial values, whereas observations show that stellar mass and gas accretion rate are correlated \citep[e.g.][]{Mohanty2005, Manara2012, Alcala2014, Alcala2017, Manara2017}. We investigate the combined effect of a lower stellar mass and a lower gas accretion rate in our detailed parameter study in Section \ref{sec:location_pressure_maximum}.

Finally, the dotted line in Fig. \ref{fig:mdot_mstar_alphadz} shows a model with a dead-zone viscosity $\alpha_{\rm DZ}=10^{-3}$ (higher than our fiducial $\alpha_{\rm DZ}=10^{-4}$). As in the simple models of \citet{Mohanty2018}, the exact value of $\alpha_{\rm DZ}$ is unimportant in the innermost, well-ionized region. In the outer regions the accretion stress is dominated by that in the dead zone, and so $\alpha_{\rm DZ}$ sets the disc structure there. This includes the disc midplane temperature, and so $\alpha_{\rm DZ}$ sets the location where, going radially inwards, the midplane temperature reaches the critical value above which the disc midplane becomes well ionized. Once again, the approximate scaling that we find, $r_{P_{\rm max}} \propto \alpha_{\rm DZ}^{-0.22}$ is close to the one expected from simple arguments laid out in Section \ref{sec:theory}. 

\subsection{Large dust grains and low accretion rates} \label{sec:big_dust_low_mdot}


Within the parameter space explored above, the steady-state solution for the inner disc structure remains qualitatively the same, with a local gas pressure maximum at the transition between the high-viscosity inner region and the low-viscosity outer region. Non-thermal sources of ionization (dominated by stellar X-rays) can suppress this picture by increasing the viscosity in the outer region and removing the pressure maximum. Within the parameter space explored above, the importance of non-thermal ionization increases for larger grains and lower gas accretion rates, but overall remains very small.

\begin{figure}
    \centering
	\includegraphics[width=0.47\textwidth,trim={0 1.5cm 0 1.5cm},clip]{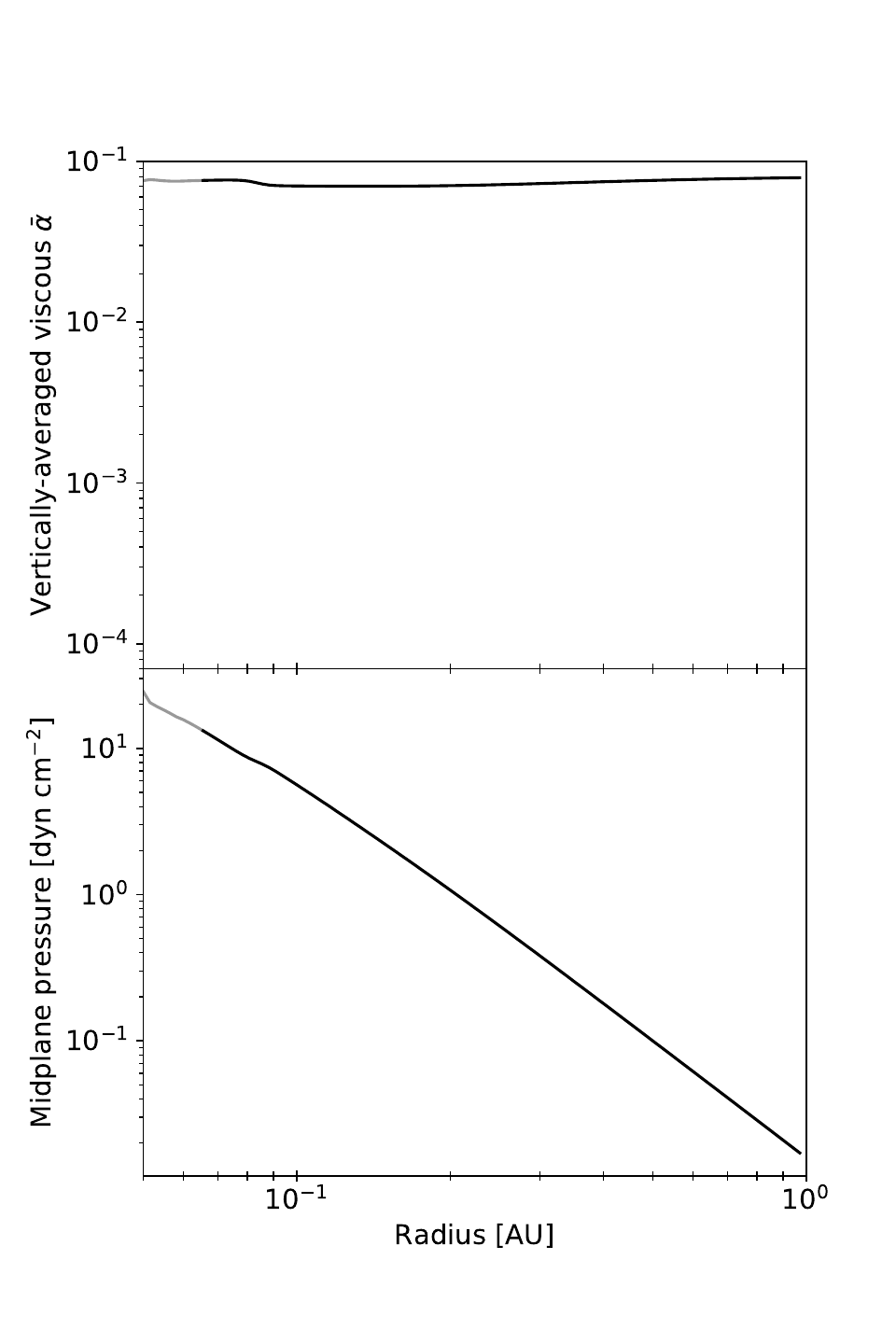}
    \caption{Vertically-averaged viscosity parameter $\bar\alpha$ (top) and midplane pressure (bottom) for a disc model with a gas accretion rate $\dot{M}=10^{-9}$\,M$_\odot$\,yr$^{-1}$, stellar mass $M_*=1$\,M$_\odot$, dead-zone viscosity $\alpha_{\rm DZ}=10^{-4}$, dust-to-gas ratio $f_{\rm dg}=10^{-2}$ and maximum grain size $a_{\rm max}=1$\,cm. There is no local gas pressure maximum in the inner disc for these model parameters. The light lines indicate the regions affected by the inner boundary condition (see Section \ref{sec:methods}). See Section \ref{sec:big_dust_low_mdot}.}
    \label{fig:fig_19}
\end{figure}

\begin{figure}
    \centering
	\includegraphics[width=0.47\textwidth,trim={0 0 0 0},clip]{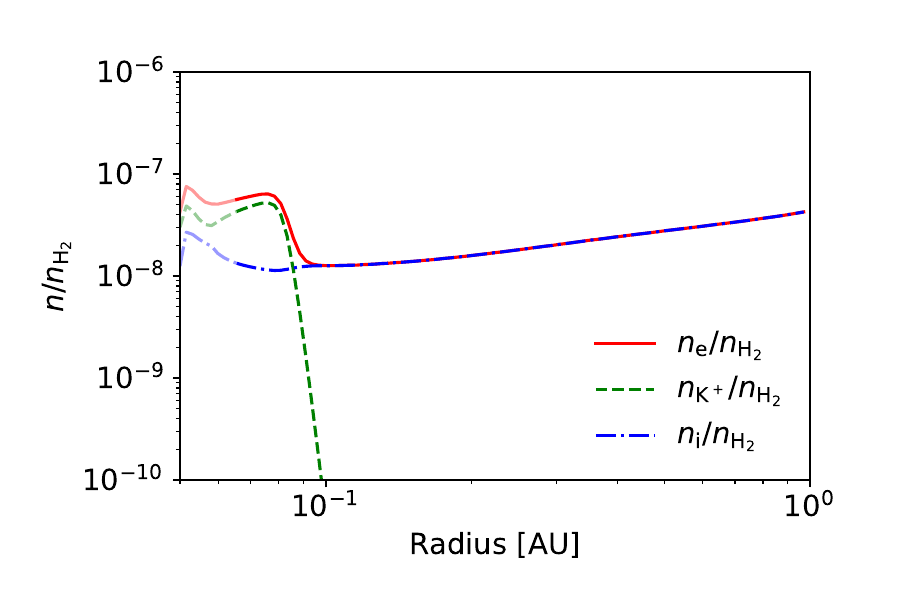}
    \caption{The fractional number densities of various ionized species at the disc midplane as a function of radius for the disc model shown in Fig. \ref{fig:fig_19}. The light lines indicate the regions affected by the inner boundary condition (see Section \ref{sec:methods}). See Section \ref{sec:big_dust_low_mdot}.}
    \label{fig:fig_19_ions}
\end{figure}

However, for large enough grains and low enough gas accretion rate, the above picture changes entirely. We find that for a maximum grain size $a_{\rm max}=1$\,cm and a gas accretion rate of $\dot{M}=10^{-9}$\,M$_\odot$\,yr$^{-1}$ (with other model parameters equal to the fiducial values) our model produces a steady-state solution in which the viscosity parameter is of the order of $\bar\alpha \sim 7 \times 10^{-2}$ throughout the inner disc, and consequently there is no local gas pressure maximum (see Fig. \ref{fig:fig_19}). Stellar X-rays are the main source of ionization in this low-surface-density solution and they drive the MRI down to the disc midplane (Fig. \ref{fig:fig_19_ions} shows that the metal ions vastly outnumber the potassium ions at the midplane). The disc is hot enough to ionise potassium inwards of 0.1\,AU and there is a drop in the midplane ionisation fraction at this radius (see Fig. \ref{fig:fig_19_ions}). However, there is no pressure maximum associated with this drop in ionisation, as the ionisation fraction remains high enough to yield a high viscosity. Moreover, the midplane ionisation fraction increases again radially outwards from this location, suggesting that the viscosity remains high in the outer disc, beyond our computational domain. Overall, these results may be interpreted as the MRI-dead zone being completely wiped out from the disc by high ionisation rate (due to low gas surface densities) and low recombination rate (due to the total surface area of the dust grains being reduced for larger maximum dust grain size).

This steady-state solution does not appear to be a unique steady-state solution for these model parameters. Multiple solutions can arise, for example, because the various thermal and non-thermal sources of ionization produce an ionisation fraction that is a complex function of temperature, density and column density (an example was discussed in Paper I). The disc ionization state is further convolved with complex MRI criteria, our assumptions about the magnetic field strength, and possibly with artefacts of the numerical procedures used (e.g. the vertical smoothing of the viscosity parameter). The uniqueness of the solution is affected both in terms of the equilibrium steady-state solution at a fixed orbital radius and a fixed value of the magnetic field strength, and in terms of the number of local maxima of the MRI-driven viscosity parameter $\bar\alpha$ as a function of the magnetic field strength. The steady-state solution shown in Fig. \ref{fig:fig_19} is favoured by our assumption that the disc structure and the magnetic field strength adapt to maximize $\bar\alpha$ at any given orbital radius. Further investigation of this issue (e.g. determining whether a steady-state solution featuring a pressure maximum can be constructed for these model parameters) is hampered by computational cost of root-solving and optimization of functions with non-unique solutions. Regardless of the existence of other solutions, the abrupt change in the steady-state structure as model parameters are varied implies that time-dependent simulations are required to examine what configuration the disc would evolve into once the X-ray-driven solution becomes viable.

Nevertheless, we can analyse under which conditions the X-ray-ionized steady-state solution can arise. This solution does not appear in a disc with small dust grains, because the higher surface area of the grains enhances the adsorption of charges onto dust grains and lowers the ionisation fraction, as discussed in previous sections. This solution also does not seem to exist at higher gas accretion rates. This can be understood as there being a maximum accretion rate that can be driven by the X-rays and the MRI, essentially due to the X-rays only ionising a limited column in the disc. 

We can estimate this maximum accretion rate as a function of model parameters and the orbital radius using a simplified model, akin to the approach of \citet{Perez-Becker2011b}. \citet{Perez-Becker2011b} focused on the role of both X-ray and UV ionisation and a much more detailed chemical network in their work, but the physics setting the maximum accretion rate is similar. First, note that the gas accretion rate is given in steady-state by $\dot{M} = 3\pi \alpha \Sigma c_s^2 / (\Omega f_r)$, where $\Sigma$ is the total gas surface density, $c_s$ is the sound speed, $\Omega$ the Keplerian velocity at a given orbital radius and $f_r = 1 - \sqrt{R_*/r}$ a factor accounting for the inner disc edge boundary condition. Second, note that this solution must feature low surface densities, so that the X-rays can ionise the disc midplane. At these low surface densities the disc is only marginally optically thick, and the disc midplane temperature is set by stellar irradiation. In fact, neglecting stellar irradiation largely suppresses the X-ray-ionized steady-state solution, since the temperature (and thus the sound speed) produced by viscous dissipation in a marginally optically thick disc is much lower than the one produced by stellar irradiation, which lowers the accretion rate given by the above expression. Therefore, in this regime, we estimate the disc temperature as $T=(\phi/2)^{1/4}(R_*/r)^{1/2}T_*$ \citep{Chiang1997}, assuming for simplicity that the incident angle of stellar irradiation $\phi$ is given by the flat-disc approximation (valid for the inner regions). As this yields a fixed radial temperature profile, at a given radius the accretion rate is maximised by maximising the product $\alpha \Sigma$.

In this regime, the MRI is primarily suppressed by ambipolar diffusion. The maximum viscosity parameter $\alpha$ that can be driven by the MRI increases with an increasing ambipolar Elsasser number Am, according to a simple function given by \citet{Bai2011a}. Assuming that the only important charged species are electrons and atomic ions and that we are in the ambipolar-diffusion-dominated regime, $\textrm{Am} \approx n_{\rm i} \langle \sigma v\rangle_{\rm i} / \Omega$, where $n_{\rm i}$ is the number density of ions and $\langle \sigma v\rangle_{\rm i}$ is the rate coefficient for momentum transfer in collisions between ions and neutrals. Therefore, the ambipolar Elsasser number is directly proportional to the number density of ions. On the other hand, in a disc ionised by stellar X-rays, the number density of ions at disc midplane decreases with increasing gas surface density, due to attenuation of the stellar X-rays. As a result, the product of the viscosity parameter and the surface density, $\alpha(\textrm{Am}(\Sigma)) \Sigma$, is maximized at some value of $\Sigma$, and so is the gas accretion rate.

Furthermore, we can estimate Am and $n_{\rm i}$ at disc midplane, by assuming the only source of ionization are X-rays and atomic ions recombine on dust grains, and using the reaction rates given in Paper I (but here we ignore the charge state of the dust grains). In our chemistry calculations we use an effective dust-to-gas ratio to mimic the grain size distribution; for the purpose of this calculation we use the effective dust-to-gas ratio that corresponds to dust grains of different sizes being weighted by their surface area (different than in the rest of our calculations), which is a more appropriate choice when dust grains only act as a recombination pathway (see section 2.4.2 in Paper I). 
We calculate the product $\alpha(\textrm{Am}(\Sigma)) \Sigma$ for a range of values of $\Sigma$ and pick the maximum value. For our fiducial stellar (Solar-mass) parameters and a maximum grain size $a_{\rm max}=1$\,cm, at the radius of 0.1\,AU we find that the maximum accretion rate that can be obtained via an X-ray-dominated solution is $\dot{M}_{\rm max} \approx 3 \times 10^{-9}$\,M$_\odot$\,yr$^{-1}$. This simple estimate thus appears to be in agreement with the results of our full numerical model.

Notably, $\dot{M}_{\rm max}$ increases with orbital radius, due to the dependence of the accretion rate on the Keplerian angular velocity, and in spite of the decrease in the X-ray ionisation rate at larger radii \citep[see also][]{Perez-Becker2011b}. Therefore, at large enough radii the X-ray-dominated steady-state solution should arise even for our fiducial accretion rate (and higher values). For the same parameters as above, within 1\,AU from the star, $\dot{M}_{\rm max}$ remains below our fiducial value of $\dot{M} = 10^{-8}$\,M$_\odot$\,yr$^{-1}$, and so the simple estimate also confirms our results from the previous sections. However, more generally, such a configuration where the disc would adopt an X-ray-dominated solution only outwards from some radius is unphysical, as it would require a sharp and large drop in the gas surface density at the transition between the different steady-state solutions. In the following section we discuss further the importance of this X-ray-dominated solution for the existence of the pressure maximum, and in Section \ref{sec:discussion} we discuss how it could fit into the general picture of disc evolution.

\section{Pressure maximum} \label{sec:location_pressure_maximum}
The above results show that, for a wide range of disc, stellar and dust parameters, an MRI accreting protoplanetary disc features a high-viscosity inner region, a low-viscosity outer region, and a gas pressure maximum at the transition between the two regions. This gas pressure maximum has been hypothesized to have a key role in the formation of the super-Earths inside the water ice line \citep{Chatterjee2014, Chatterjee2015, Hu2016, Hu2018}. In this section, we wish to examine in more detail the existence and location of the pressure maximum.

As discussed in Section \ref{sec:theory}, the existence of the pressure maximum in the inner disc requires two conditions. First, the ionization fraction should increase (sharply) above some critical temperature. Second, the disc should be hot enough so that the ionization fraction in the innermost region increases above a value required to sustain the MRI. The first of these conditions is not fulfilled if the disc evolves into a steady-state structure primarily ionized by stellar X-rays, discussed in Section \ref{sec:big_dust_low_mdot}. As noted in Section \ref{sec:big_dust_low_mdot}, time-dependent simulations are required to establish whether the stellar X-rays could indeed overtake the disc structure and at what point in the disc evolution this could happen. Still, it is important to consider this possibility when discussing the existence of the pressure maximum in the inner disc. Due to computational cost, it is not possible to calculate a large grid of models which include the possibility of multiple equilibrium solutions, as well as heating by stellar irradiation (which is, as discussed above, key for the existence of X-ray-dominated solutions). Therefore, we proceed as follows.

In this section, we neglect the heating by stellar irradiation and the ionization of the disc by the stellar X-rays, and use our model to find the radial location of the pressure maximum for various combinations of disc and stellar parameters. Then, for each combination of parameters, we estimate the maximum gas accretion rate that can be attained in an irradiated X-ray-ionized disc, at that radial location, using the simplified model outlined in Section \ref{sec:big_dust_low_mdot}. If this maximum gas accretion rate of an X-ray-ionized solution is larger than the given, input accretion rate, we note that the pressure maximum \textit{might} not exist for those parameters. This condition for the existence of the pressure maximum is likely too strict. The maximum accretion rate of an X-ray-dominated solution increases with increasing orbital radius, and so the above condition does not guarantee that this solution would exist all the way to the inner disc edge for the given parameters. Nevertheless, this condition can also be understood as an estimate of whether a surface X-ray-ionized layer may become more important than the accretion through an MRI-dead midplane, i.e., whether the X-rays could perturb the steady-state solution featuring a pressure maximum. Ultimately, time-dependent simulation and a more complex chemical network will be needed to investigate how a real disc would behave, and we discuss this further in Section \ref{sec:limitations}. 

\begin{figure}
    \centering
	\includegraphics[width=0.49\textwidth,trim={0 0 0 0},clip]{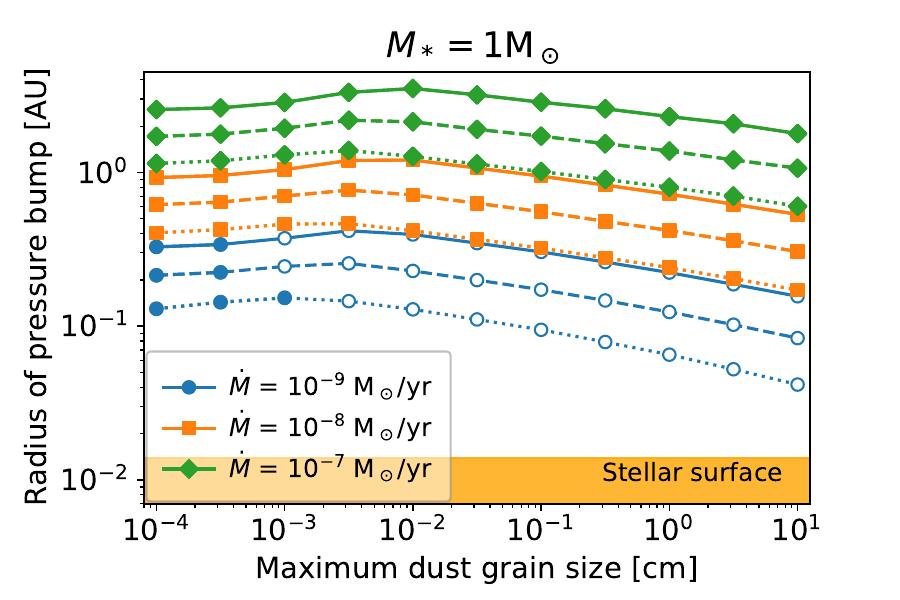}
	\includegraphics[width=0.49\textwidth,trim={0 0 0 0},clip]{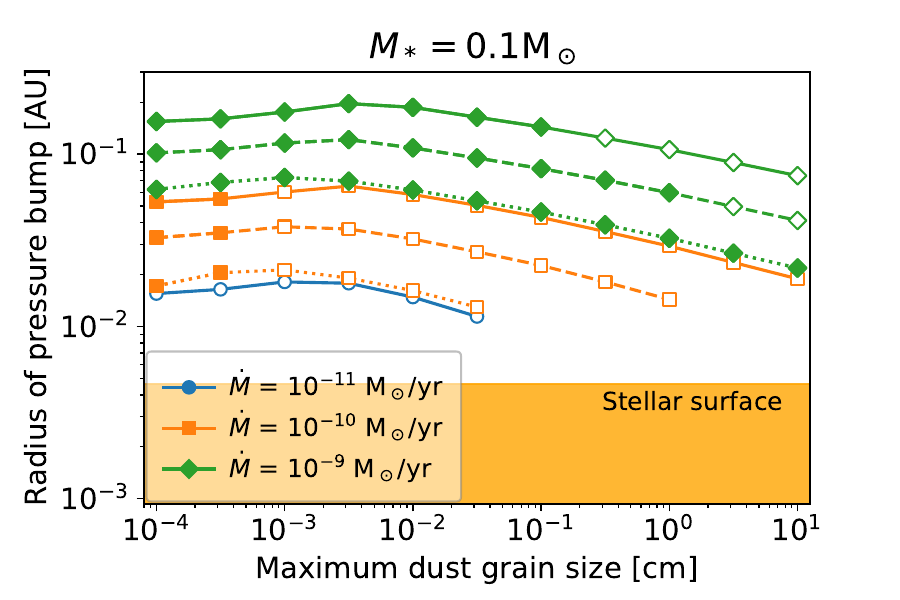}
    \caption{Radius of gas pressure bump as a function of maximum dust grain size. Top panel shows the results for a stellar mass $M_*=1$\,M$_\odot$, bottom panel for $M_*=0.1$\,M$_\odot$. In all models the dust-to-gas ratio is $10^{-2}$. Solid, dashed and dotted lines show results for the dead-zone viscosity parameter $\alpha_{\rm DZ}=10^{-5}$, $10^{-4}$ and $10^{-3}$, respectively. Blue, green and red lines indicate different gas accretion rates $\dot{M}$ as indicated in plot legends for each panel. To obtain these results, we neglect stellar irradiation in our disc model, and use a simple calculation to estimate if including the irradiation would produce a steady-state solution ionized primarily by stellar X-rays and featuring no pressure maximum (shown as empty symbols). This is indeed found to be the case for lower accretion rates and larger dust grains. For $M_*=0.1$\,M$_\odot$, there may be no pressure maximum for the lower end of the observed accretion rates ($\dot{M}=10^{-11}$\,M$_\odot$\,yr$^{-1}$) for any maximum grain size. See Section \ref{sec:location_pressure_maximum}.}
    \label{fig:rpmax_adust}
\end{figure}

Fig. \ref{fig:rpmax_adust} shows the radial location of the pressure maximum (or pressure bump) as a function of the maximum dust grain size (see Section \ref{sec:methods} for details of the dust grain distribution and how it is mimicked in our chemical network), for a stellar mass of $1$\,M$_\odot$ (the top panel) and $0.1$\,M$_\odot$ (the bottom panel). Those models in which the X-rays \textit{may} erase the pressure maximum are indicated by empty symbols. In all models shown here, we adopt our fiducial dust-to-gas ratio of $10^{-2}$. The solid, dashed and dotted lines correspond to different values of the dead-zone viscosity parameter ($\alpha_{\rm DZ}=10^{-5}$, $10^{-4}$ and $10^{-3}$, respectively). For the different stellar masses we explore different ranges of the gas accretion rate $\dot{M}$, as indicated in plot legends next to each panel. The chosen ranges are motivated by observational studies, which find that for a Solar-mass star typically $\dot{M} \sim 10^{-8}$\,M$_\odot$\,yr$^{-1}$ and for the stellar mass of $0.1$\,M$_\odot$, typically $\dot{M} \sim 10^{-10}$\,M$_\odot$\,yr$^{-1}$ \citep[e.g.][]{Mohanty2005, Manara2012, Alcala2014, Alcala2017, Manara2017}. There is a significant spread both in the reported mean values in these studies ($\pm 1$\,dex for the stellar mass of $0.1$\,M$_\odot$, and somewhat less for a Solar-mass star) and within the observed samples in each study (up to $2$\,dex). However, the correlation with the stellar mass appears robust, and so we adopt the above typical values as mean values and vary the gas accretion by $\pm 1$\,dex for each stellar mass.

For the Solar-mass star we find that, for the upper end of the observed range of gas accretion rates ($\dot{M} \gtrsim 10^{-8}$\,M$_\odot$\,yr$^{-1}$) and a wide range of $\alpha_{\rm DZ}$, the existence of the pressure maximum is robust for a wide range of grain sizes. For the lower gas accretion rate ($\dot{M} = 10^{-9}$\,M$_\odot$\,yr$^{-1}$) and grain sizes larger than a few microns, the simple estimate described above suggests that accretion through an X-ray-ionized surface layer may perturb the disc structure. Therefore, we caution that a pressure maximum may not always occur for these parameters. For $M_*=0.1$\,M$_\odot$ the portion of the parameter space within which the pressure maximum is always present is even smaller, since the observed mean gas accretion rate is two orders of magnitude lower. Note that the range of grain sizes for which the pressure maximum is unperturbed for $\dot{M} = 10^{-9}$\,M$_\odot$\,yr$^{-1}$ around the lower-mass star is larger than for the Solar mass star, due to the lower-mass star having a lower (X-ray) luminosity. 

The radius of the pressure maximum as a function of the maximum dust grain size ($a_{\rm max}$) shows a similar trend across the various values of the stellar mass, accretion rate and dead-zone viscosity: it weakly increases with increasing $a_{\rm max}$ for small grains, peaks at about $a_{\rm max} \sim 10^{-2}$\,cm, and then steadily decreases for larger dust grains. The factors causing this have already been briefly discussed in section \ref{sec:ratio_size}. First, recall that an increase in the disc opacity means that accretion heat can escape less easily, making the disc midplane hotter and pushing the pressure maximum outwards. In addition, for small dust grains, larger dust grain size leads to a moderate increase in the opacity (here, the relevant opacity is the opacity of the disc to its own radiation in the optically-thick regions, i.e., the Rosseland-mean opacity, see Fig. \ref{fig:opacities}). At the same time, the increase in dust grain size reduces the critical temperature at which ionization fraction rises due to thermionic and ion emission \citep[as the increase in dust grain size is equivalent to a reduction in the effective dust-to-gas ratio in our chemical network; see also][]{Desch2015}, pushing the pressure maximum outwards. When these factors are compounded, for small grains, the radius of the pressure maximum increases with $a_{\rm max}$. However, if the grains grow beyond $a_{\rm max} \sim 10^{-2}$\,cm, dust opacities are severely decreased with increasing grain size and the net effect is a decrease in the radius of the pressure maximum. Additionally, note that the exact value of $a_{\rm max}$ at which the radius of the pressure bump peaks varies somewhat with $\dot{M}$ and $\alpha_{\rm DZ}$; this can be expected, since both determine the gas surface density and thus the optical depth at disc midplane and the relative importance of the above two effects. It is also important to note that including heating due to stellar irradiation seems to somewhat modify the trend for small dust grains, as can be seen by comparing Fig~\ref{fig:rpmax_adust} to the models discussed in the previous section. Specifically, Fig. \ref{fig:chem_ratio_size} shows that for our fiducial disc and stellar parameters, micron-size grains result in a pressure maximum at a larger radius. The resulting differences in the radial location of the pressure maximum are, however, small, and certainly for larger grains the pressure maximum moves inwards.

\section{Discussion} \label{sec:discussion}
In this work we have investigated how the structure of the inner disc, accreting primarily through the MRI, changes with various disc, stellar and dust parameters. Of particular interest are the existence and the location of a local gas pressure maximum and a highly-turbulent region inwards of it, which could accumulate dust grains drifting in radially from the outer disc, possibly leading to the formation of planetary cores \citep{Chatterjee2014, Hu2018, Jankovic2019}. The models presented in this work are steady-state models, each with a distribution of dust grains that is fixed throughout the disc. However, as discussed in Sections \ref{sec:disc_dust_growth} and \ref{sec:dust_accumulation}, these models provide us with important insights into how the inner disc could evolve as the dust grains grow, if and how the dust will accumulate, how this accumulation could feedback on the gas structure, and the disc parameters that are favourable for the formation of planetary cores. In Section \ref{sec:limitations} we discuss the various limitations of these models.

Before we proceed, we re-iterate that dust is incredibly important for the inner disc structure. Thermionic and ion emission from dust grains and the adsorption of charges from the gas onto the dust grains control the ionization fraction, and thus the extent of the inner region where the MRI can drive efficient accretion. These processes yield an ionization fraction that increases sharply above a critical temperature, similar to thermal ionization. This critical temperature appears to be a slowly varying function of dust properties, however, and so it is the effect of the dust on the disc opacity that largely determines the location of the pressure maximum. Increasing the dust opacity results in a more optically-thick, hotter inner disc, with a radius of the pressure maximum at a larger distance from the star.

\subsection{Dust growth} \label{sec:disc_dust_growth}
As discussed in Section~\ref{sec:location_pressure_maximum}, dust growth to $a_{\rm max} \sim 10^{-2}$\,cm increases the extent of the high-viscosity inner region and the radius at which the pressure maximum is located, as an increase in dust grain size leads to a moderate increase in the disc opacity and a decrease in the threshold temperature at which thermionic and ion emission become efficient. Growth beyond that size has the opposite effect, as it leads to a significant decrease in the disc opacity, making the disc midplane colder, and thus less ionized. Therefore, in the inner disc, if dust grows larger than $\sim 100$\,$\mu$m sizes, the dead-zone inner edge moves inwards.

Note that this is the opposite of what happens in the outer regions of protoplanetary discs. The outer regions are ionized primarily by the stellar X-rays and cosmic rays. These sources of ionization become more important further away from the star, as the disc column density decreases and high-temperature effects become unimportant. These regions are expected to be optically thin to their own radiation, and the primary source of heat is stellar irradiation. Therefore, the dust acts primarily to lower the ionization fraction by adsorbing free charges from the gas. Because of this, in the outer regions the dead zone is expected to shrink as the dust grains grow \citep{Sano2000, Ilgner2006}.

We can calculate the location of the pressure maximum under an assumption that the maximum dust grain size has reached a growth limit. In the inner disc, dust growth is limited by collisional fragmentation of dust grains due to relative turbulent velocities \citep{Birnstiel2010, Birnstiel2012, Drazkowska2016}. Here, the relative grain velocities are induced by the MRI-driven turbulence, as well as the lower levels of turbulence assumed to persist in the MRI-dead zone. Since this growth limit depends on the velocities of dust grains due to turbulent velocities of the gas, it is given in terms of the particles ``Stokes number'' ($\textrm{St}$), the ratio between the particle gas drag stopping time and the eddy turnover time \citep[where the eddy turnover time is taken to be $1/\Omega$][]{Zhu2015}. In a turbulent disc, typical collisional relative velocity between dust grains is given by $V_{\rm dd}^2 \approx 3 V_{\rm g}^2 \textrm{St}$ \citep[for St < 1,][]{Ormel2007}, where $V_{\rm g}$ is the typical turbulent gas velocity (given by $V_{\rm g}^2=\alpha c_{\rm s}^2$). There is a critical velocity $u_{\rm frag}$ above which a collision between dust grains results in their fragmentation, rather than sticking/growth. For silicate grains of similar size, $u_{\rm frag} \sim 1$\,m\,s$^{-1}$ \citep[][although note that grains might become more sticky at the high temperatures present in the inner disc \citep{Demirci2019}]{BlumMunch1993, Beitz2011, Schrapler2012, Bukhari_Syed2017}. Since the Stokes number St is directly related to the grain size, and the collision velocity to St, fragmentation imposes an upper limit on dust growth. At the fragmentation limit \citep{Birnstiel2009, Birnstiel2012}, 
\begin{equation} \label{eq:frag_limit}
    \textrm{St}_{\rm frag} = \frac{ u_{\rm frag}^2 }{ 3\alpha c_{\rm s}^2 }.
\end{equation}

The exact relationship between the Stokes number and the particle size depends on the relevant drag law \citep{Weidenschilling1977}. Typically, the dust grains in protoplanetary discs are smaller than the mean free path of gas molecules, and therefore couple to the gas according to the Epstein drag law. However, due to the high densities in the inner disc, dust grains may enter the Stokes regime. Importantly, the above approximate expression for the turbulent relative velocity between dust grains ($V_{\rm dd}$) has been derived under an assumption that St does not depend on the relative velocity between the dust grain and the gas, $V_{\rm dg}$. This assumption is true for grains in the Epstein drag regime. In the Stokes regime, it is true only if the Reynolds number Re of the particle is less than unity. We always check that this condition is fulfilled for our particles in the Stokes regime, and that we may employ the above expression for $V_{\rm dd}$. The Reynolds number of a particle itself depends on the velocity $V_{\rm dg}$, for which we adopt another approximate expression, $V_{\rm dg}^2 = V_{\rm g}^2 \textrm{St}/(1+\textrm{St})$ \citep[][note that this expression was derived analytically for $\textrm{St} \ll 1$, but also shown to be applicable for a wide range of St through a comparison with numerical simulations]{Cuzzi2003}.

To calculate the location of the pressure maximum in a disc in which grain growth is limited by fragmentation, we proceed as follows. At the location of the pressure maximum for various combinations of stellar mass, accretion rate, dead-zone viscosity parameter and maximum dust grain size (i.e., for every point in Fig. \ref{fig:rpmax_adust}), we calculate the fragmentation limit for the particle Stokes number, St$_{\rm frag}$ (assuming $\alpha=\bar\alpha$), and the corresponding grain size, $a_{\rm frag}$ (for an appropriate drag law). For each combination of the disc and stellar parameters (stellar mass, accretion rate and dead-zone viscosity parameter), this yields a set of points describing a function $a_{\rm frag}(a_{\rm max})$. We connect these points using linear interpolation, and find the maximum grain size such that $a_{\rm frag}(a_{\rm max})=a_{\rm max}$. 
Then, again using the results shown in Fig. \ref{fig:rpmax_adust} and linear interpolation, we find the radius of the pressure maximum at that value of $a_{\rm max}$. Similarly, we find the corresponding midplane temperature and density.

This calculation utilizes models in which the maximum dust grain size is assumed to be constant everywhere in the disc, and so the obtained solutions also formally correspond to models in which the maximum dust grain size is radially constant (and equal to the fragmentation limit $a_{\rm frag}$ at the pressure maximum). In a real disc, the fragmentation limit to which particles can grow would be a function of the turbulence levels and other parameters which vary as functions of radius. While this calculation does not take this radial variation of dust size into account, the fragmentation limit at the pressure maximum and the location of the pressure maximum would remain the same as in the solutions found here. In particular, note that radially inwards from the pressure maximum, $a_{\rm frag}$ should decrease compared to the value at the pressure maximum, as the turbulence parameter $\alpha$ increases. 
Furthermore, for large maximum grain sizes $a_{\rm max}$ (which is the regime pertaining to the solutions found here), a decrease in $a_{\rm max}$ yields an increase in $\alpha$ at a fixed radius. Thus, if we accounted for the decrease in $a_{\rm max}=a_{\rm frag}$ inwards of the pressure maximum, this would only make the radial gradient of $\alpha$ steeper inwards of the pressure maximum, but it would not change the location of its minimum, and therefore not the location of the pressure maximum obtained here.

\begin{figure}
    \centering
	\includegraphics[width=0.47\textwidth,trim={0 3cm 0 4cm},clip]{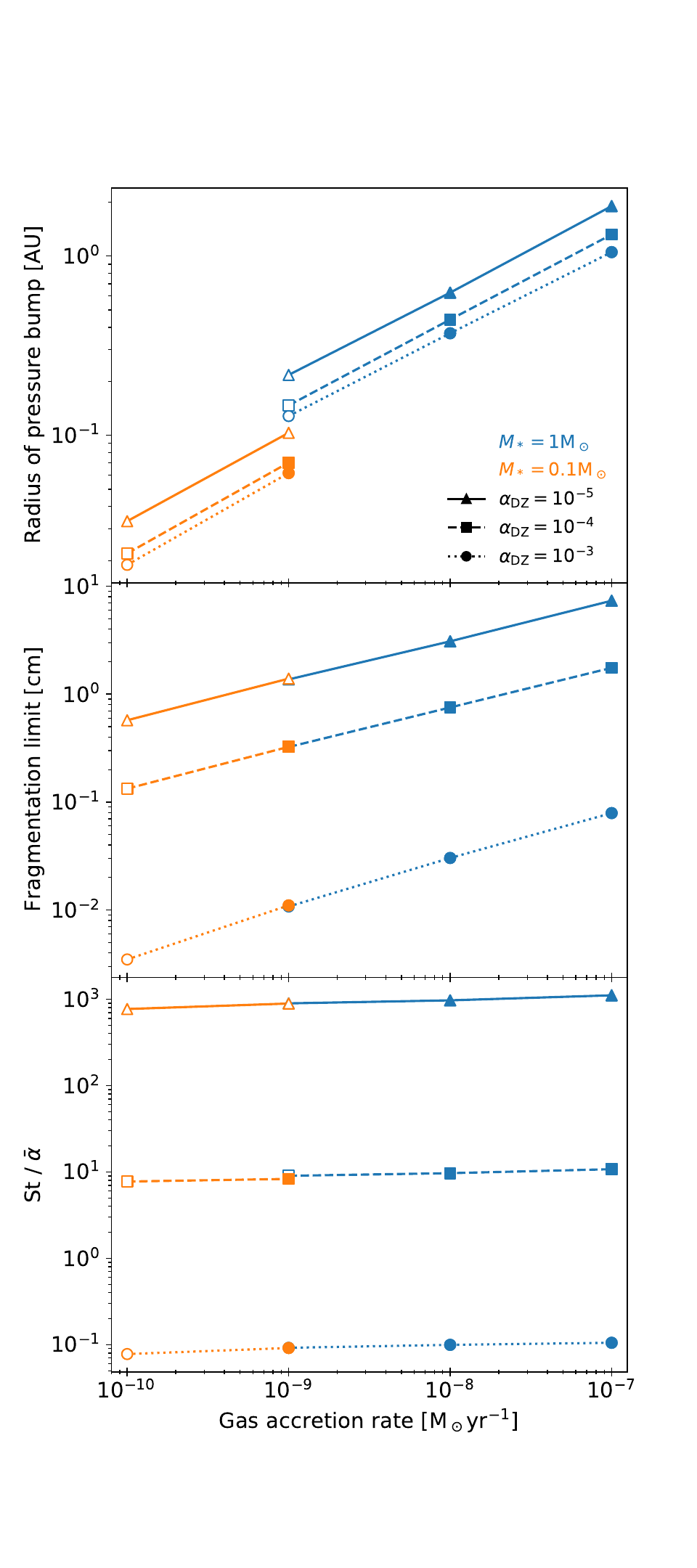}
    \caption{Radius of the pressure bump (top), maximum dust grain size (middle) and ratio of the Stokes number to the viscosity parameter ($\textrm{St}/\bar\alpha$) at the pressure bump (bottom) as functions of the gas accretion rate, for a maximum dust grain size that corresponds to the grain growth limit due to turbulent fragmentation. Solid, dashed and dotted lines correspond to different values of the dead-zone viscosity parameter $\alpha_{\rm DZ}$ as shown in the plot legend. The dust grain size and $\textrm{St}/\bar\alpha$ are highly sensitive to the dead-zone viscosity parameter $\alpha_{\rm DZ}$. Conversely, varying $\alpha_{\rm DZ}$ affects the location of the pressure bump only weakly. Blue lines show the results for a stellar mass $M_*=1$\,M$_\odot$, orange lines for $M_*=0.1$\,M$_\odot$. Empty symbols denote solutions where an alternative steady-state solution may exist that is ionized primarily by stellar X-rays and featuring no pressure maximum, same as in Fig. \ref{fig:rpmax_adust}. There is no solution for $M_*=0.1$\,M$_\odot$ and $\dot{M} = 10^{-11}$\,M$_\odot$\,yr$^{-1}$.
    See Sections \ref{sec:disc_dust_growth} and \ref{sec:dust_accumulation}.}
    \label{fig:rpmax_mdot}
\end{figure}

The results for the radius of the pressure maximum and the grain size are shown in Fig. \ref{fig:rpmax_mdot}, as functions of the gas accretion rate, for different values of stellar mass and the dead-zone viscosity parameter. The maximum grain size at the pressure maximum is limited by turbulent fragmentation (middle panel) and thus sensitive to the dead-zone viscosity parameter $\alpha_{\rm DZ}$ (results for $\alpha_{\rm DZ}=10^{-5}$, $10^{-4}$ and $10^{-3}$ are shown by the solid, dashed and dotted lines, respectively). This is because, in these models, at the location of the pressure maximum, the vertically-averaged viscosity (and turbulence) parameter $\bar\alpha=\alpha_{\rm DZ}$.

Remarkably, despite the sensitivity of the grain size to $\alpha_{\rm DZ}$, the radius of the pressure maximum depends very weakly on this parameter. 
This is a result of three different inter-dependencies. At a fixed maximum grain size (and other parameters), a lower $\alpha_{\rm DZ}$ yields a larger radius of the pressure bump, and also a larger fragmentation-limited grain size. Concurrently, a larger maximum grain size yields a smaller radius of the pressure bump (for maximum dust grain size larger than about $10^{-2}$\,cm, see Fig. \ref{fig:rpmax_adust}). Evidently, compounding these inter-dependencies results in a weakly-varying radius of the pressure bump. 

We can also compare these results against simple theoretical expectations, by coupling eq. (\ref{eq:ss_pmax_dust}) with the expressions for the fragmentation-limited dust grain size (obtained from eq. (\ref{eq:frag_limit}) and the appropriate drag laws). In the Epstein drag regime (pertaining to the solution for $\alpha_{\rm DZ} = 10^{-3}$ in Fig. \ref{fig:rpmax_mdot}), we obtain for the radius of the pressure bump
\begin{equation}
     r_{P_{\rm max}}{\rm \,\,}\propto{\rm \,\,} f_{\rm dg}^{4/15} \dot{M}^{2/5} M_*^{1/3}
     {\rm \,\,}\propto{\rm \,\,} f_{\rm dg}^{0.27} \dot{M}^{0.4} M_*^{0.33} ,
\end{equation}
and in the Stokes regime (pertaining to the solutions for $\alpha_{\rm DZ} \leq 10^{-4}$),
\begin{equation}
     r_{P_{\rm max}}{\rm \,\,}\propto{\rm \,\,} f_{\rm dg}^{8/39} \alpha^{-2/13} \dot{M}^{16/39} M_*^{1/3}
     {\rm \,\,}\propto{\rm \,\,}f_{\rm dg}^{0.2} \alpha^{-0.15} \dot{M}^{0.41} M_*^{0.33} .
\end{equation}
These simple predictions agree with our finding that the dependency of the radius of the pressure bump on $\alpha_{\rm DZ}$ is very weak, and in fact expected to be non-existent in the Epstein drag regime. The dependencies on the other parameters also seem to match fairly well. Small deviations are to be expected since in the numerical models the pressure bump does not occur precisely at a constant fixed temperature.

Furthermore, for the Solar-mass star, the pressure maximum is located at radii between $\sim 0.3$\,AU and $\sim 2$\, AU in the upper end of the range of the observationally-motivated gas accretion rates considered here. However, for the accretion rate of $\dot{M} = 10^{-9}$\,M$_\odot$\,yr$^{-1}$ the pressure maximum may be perturbed and possibly removed from the inner disc altogether. As discussed in Section \ref{sec:location_pressure_maximum}, for lower gas accretion rates and larger dust grain sizes, we find that the disc could assume a high-viscosity X-ray-dominated steady-state solution which features no pressure maximum (in Fig. \ref{fig:rpmax_mdot}, this is indicated by empty symbols).

For a lower-mass star, $M_*=0.1$\,M$_\odot$, at a fixed gas accretion rate, dust grains grow to similar sizes as for the Solar-mass star. 
Overall, the radius of the pressure maximum is expected to be smaller, due to the lower viscous dissipation rate and the lower observed gas accretion rates (see the discussions in sections \ref{sec:mdot_mstar_alphadz} and \ref{sec:location_pressure_maximum}). Analogously to the case of the Solar-mass star, there might not be a pressure maximum at lower accretion rates.  

\subsection{Dust accumulation} \label{sec:dust_accumulation}
For dust grains to become trapped within the pressure maximum, the outwards radial drift velocity of dust grains just inwards of the pressure maximum should be higher than the velocity with which the accreting gas advects the grains inwards. The ratio of the radial drift and the gas advection velocities is roughly equal to the ratio of the particle Stokes number and the viscous $\alpha$ \citep{Jacquet2012}. Therefore, for the particle radial drift to overcome advection with the gas inwards of the pressure maximum, it is required that $\textrm{St}/\alpha > 1$. To check whether this condition is fulfilled, in the bottom panel of Fig. \ref{fig:rpmax_mdot} we show the ratio between the Stokes number at the fragmentation limit $\textrm{St}_{\rm frag}$ and the vertically averaged viscosity and turbulence parameter $\bar\alpha$ at the location of the pressure maximum.

The ratio $\textrm{St}_{\rm frag}/\bar\alpha$ is most sensitive to the value of the dead-zone viscosity parameter $\alpha_{\rm DZ}$. The Stokes number at the fragmentation limit $\textrm{St}_{\rm frag}$ is given by eq. (\ref{eq:frag_limit}). At a fixed critical fragmentation velocity $u_{\rm frag}$, it is a function only of the viscosity parameter $\bar\alpha$ and the speed of sound $c_{\rm s}$ (i.e., the temperature) at the pressure maximum. Since these solutions are extracted from models which neglect ionization by stellar X-rays, $\bar\alpha=\alpha_{\rm DZ}$, the value of the dead-zone viscosity parameter. Hence, $\textrm{St}_{\rm frag}/\bar\alpha\, \propto\, 1/\alpha_{\rm DZ}^{2}$. This ratio also varies slightly with gas accretion rate, which we can attribute to the fact that the pressure maximum does not occur precisely at a fixed critical temperature. Instead, the ionization fraction, driven primarily by thermionic and ion emission, also depends on the dust properties and the disc density.

Overall, whether the dust grains become trapped in the pressure maximum depends on an assumed value of $\alpha_{\rm DZ}$. We may consider how the disc might evolve forward, depending on this value. First, if the dead-zone viscosity parameter is low, $\alpha_{\rm DZ}=10^{-5}$, $\textrm{St}_{\rm frag}/\bar\alpha \gg 1$. In this case, dust grains could readily accumulate at the pressure maximum.  This accumulation might lead to an unstable configuration, as an increase in the dust-to-gas ratio leads to an increase in $\bar\alpha$ at a given radius (see Fig. \ref{fig:chem_ratio_size}). As the dust-to-gas ratio would increase at the pressure maximum, and decrease either side, this could lead to an emergence of an additional minimum in $\bar\alpha$ outwards from the original minimum, and thus to a formation of an additional pressure trap. Time-dependent simulations are needed to examine further evolution of the disc. In general, if the pressure maximum efficiently traps the radially-drifting dust, one observable consequence would be that the gas accreting onto the star would be depleted in elements of which the dust is composed: such a mechanism has been proposed to be at work in the inner disc of TW Hya, where the levels of depletion of refractory and volatile elements suggest a dust trap inside of the water ice line \citep{McClure2020}, and also in transition discs surrounding Herbig Ae/Be stars \citep{Kama2015}.

Second, consider a case in which $\textrm{St}_{\rm frag}/\bar\alpha \sim 1$, which would occur if the dead-zone viscosity parameter is slightly above the middle of the plausible range, $\alpha_{\rm DZ}=10^{-4}$. In this case, considered by \citet{Jankovic2019}, the gas pressure maximum does not trap large amounts of dust. This is because $\bar\alpha$ increases inwards of the pressure maximum, and $\textrm{St}_{\rm frag}$ decreases. This limits the radial width of the pressure trap. Dust advection with the accreting gas is not sufficient to remove the dust grains from the trap; however, the grains are also mixed radially by the turbulence. The dust-to-gas ratio at the pressure maximum is then limited by the corresponding radial diffusion term. 
Nevertheless, dust would still accumulate in the entire region interior to the pressure maximum, as in the highly-turbulent innermost region dust grains become small enough to couple to the gas, reducing the radial drift relative to the outer disc. However, in this case, the amount of dust that can be accumulated in the inner disc is limited by the flux of dust drifting inwards from the outer disc. Moreover, the small size of the dust grains is detrimental to planet formation.

In this work, we also find that a higher dust-to-gas ratio yields a larger extent of the high-viscosity inner region (see Section \ref{sec:chem_ratio_size}). This is highly beneficial for planet formation in the inner disc, as it implies that the accumulation of dust is not only sustainable, but also leads to a radial expansion of the high-viscosity, high-turbulence region inside of which the dust accumulates. In particular, this expansion is beneficial for the growth of the small fragmentation-limited dust grains into larger, more rigid solid bodies, i.e., into planetesimals. Specifically, planetesimals may form out of dust grains through a combination of the streaming instabilities (SI) and the gravitational instability \citep[][]{Youdin2005, Johansen2007, Bai2010, Johansen2012, Simon2016, Simon2017, Schafer2017}. Under certain conditions, the SI leads to localized concentrations of dust grains susceptible to gravitational collapse. \citet{Jankovic2019} pointed out that this process is unlikely in the inner disc if the pressure (and the density) maximum is located at very short orbital distances, as too close to the star the tidal effect of the star prevents the gravitational collapse. Therefore, the shift of the pressure maximum to larger orbital distances due to the accumulation of dust could potentially help to overcome this barrier and form planetesimals. 

Finally, if the dead-zone viscosity parameter is high, $\alpha_{\rm DZ}=10^{-3}$, $\textrm{St}_{\rm frag}/\bar\alpha \ll 1$. In this case, dust grains are so small and well-coupled to the gas that they are advected through the pressure maximum inwards. Dust may still accumulate interior to the pressure maximum, as a consequence of fragmentation in the innermost regions, as noted above. However, it is unlikely that this could lead to the formation of larger solid bodies. While the exact value of the dead-zone viscosity is unimportant for the location of the pressure maximum (see Fig. \ref{fig:rpmax_mdot} and Section \ref{sec:disc_dust_growth}), in this case the grains would likely be too small, too well-coupled to the gas to start the streaming instabilities \citep{Carrera2015, Yang2017}.

\subsection{Limitations} \label{sec:limitations}
In this work, we have assumed that the MRI-accreting inner disc is in an equilibrium, steady state. However, there are several processes whose further study requires to consider the time dependence of the disc structure.

First, even if variations in the dust properties are neglected, the steady-state MRI-accreting inner disc is unstable to surface density perturbations \citep{Mohanty2018}. At a fixed radius, the MRI-driven accretion rate decreases with an increasing surface density (as calculated using the steady-state models). Therefore, a perturbation in the disc surface density might lead out of the steady-state, creating a pile up of mass in a certain region. This so-called viscous instability \citep{Lightman1974, Pringle1981} could have important consequences for the inner disc structure and planet formation, as it is likely to produce rings and gaps on the viscous timescale.

Second, the results presented in this work show that the accumulation of dust in the inner disc is possibly a highly dynamical process. The dust grain size and the dust-to-gas ratio strongly affect the MRI-driven viscosity at a given radius, and, concurrently, the extent of the innermost region within which the dust may accumulate. The evolution of the disc is particularly unclear if the dust is trapped in the pressure maximum, e.g., whether further evolution of gas and dust might lead to a modified equilibrium state, or creation of multiple pressure maxima.

Third, while we considered the effects of dust on the disc thermal and chemical structure, we did not account for the dynamical effects. If dust grains drift radially due to gas drag, there is also a back-reaction on the gas \citep{Nakagawa1986}. This becomes increasingly important at high dust-to-gas ratios. In particular, drag back-reaction acts to flatten the radial gas pressure profile, and so it would affect accumulation of dust in the pressure maximum \citep{Taki2016}. Further study of the early stages of planet formation in the inner disc should account for the self-consistent, time-dependent evolution of the gas and the dust.

Fourth, it was shown that for some disc parameters dust growth could lead to steady-state solutions in which the disc is primarily ionized by stellar X-rays and no pressure maximum would exist. This deserves further study through time-dependent simulations, and also using a more detailed chemical network than considered here. In the regime in which the high-temperature effects are unimportant, at low dust-to-gas ratios (equivalent to larger grain sizes) simplified chemical networks (such as the one used here) overestimate the disc ionization fraction, and the MRI-driven accretion efficiency, compared to the more complex networks \citep{Ilgner2006}. The stellar X-ray luminosity may also need to be more carefully considered, as the stellar parameters used here are adopted from very early times in the stellar evolution models, while low gas accretion rates (for which the X-ray-dominated solutions appear) are observed at later stages of protoplanetary disc evolution. 
Lastly, propagation of the X-rays may need to be treated more accurately at short periods, and the penetration of X-rays from the ``bottom'' side of the disc should be taken into account when X-rays can reach the disc midplane.

Furthermore, we showed that the inner disc structure is sensitive to the disc opacities. Yet the radiative properties of the disc are determined only by silicate dust grains in this work. Other important species \citep[e.g. carbonaceous grains;][]{Pollack1994} could condense in some of the colder regions of the disc (e.g. near the disc photosphere - below the upper layers heated by stellar irradiation and above the hot, optically-thick disc midplane), and their contribution to the opacities could alter the details of the location of the pressure maximum. Opacities due to atomic and molecular lines have also been neglected. In the optically-thick regions (such as the disc midplane in our models), this is a good assumption since the Rosseland-mean opacity of the gas is always negligible compared to that of the dust where the pressure maximum is located. However, in the optically-thin regions (in the disc upper layers), Planck-mean opacity of the gas at high temperatures is comparable to that of micron-sized dust, and the absorption coefficient $\kappa_{\rm P}^*$ greatly exceeds that of cm-sized dust \citep[e.g.][]{Malygin2014}. Since the gas accretes primarily through the dense, optically-thick regions around the disc midplane, we can expect that including the gas opacities would not change our results, as the higher absorption of stellar light would only increase the temperature in the uppermost disc layers. Nevertheless, note that the gas opacities are strongly non-monotonic, and the upper disc layers might not be in thermal equilibrium \citep{Malygin2014}, which is assumed to be the case here. 

Another limitation of this work in modelling of the disc upper layers is that the possibility of shadowing neglected. As in other 1+1D models \citep[e.g.][]{Chiang1997,DAlessio1998}, when heating by stellar irradiation is considered, there is an underlying assumption that the disc is flaring, so that stellar rays can reach the disc surface at all radii. However, \citet{Terquem2008} showed that the inner boundary of the MRI-dead zone may be puffed up sufficiently to throw a shadow over the outer regions of the disc. In our models, we can look for this possibility by inspecting the irradiated surface of the disc arising from the integration of the disc structure in the vertical direction \citep[as opposed to the adopted irradiated surface, constructed by ray-tracing in two dimensions under the assumption that the disc is flaring, see Paper I;][]{DAlessio1999}. Indeed, we find that, in the vicinity of the pressure maximum there is an inconsistency between the two surfaces, with the former surface implying that the region immediately outwards from the pressure maximum should be in a shadow. Nonetheless, we do not expect that correcting for the shadowing would change any of the conclusions of this work due to the same reason as above: temperature at the disc midplane is primarily determined by viscous dissipation, and not by stellar irradiation.


Furthermore, in this work we have assumed that the disc accretes viscously, via the MRI, and that the resulting viscosity is well described by criteria extracted from local magnetohydrodynamic simulations. Accretion in protoplanetary discs may also be driven by non-viscous processes, e.g. by large scale laminar flows if Hall effect is the dominant non-ideal MHD effect \citep[][]{Lesur2014}, or, in the presence of a magnetic field threading the disc, by magnetic winds \citep[][]{Suzuki2009, Suzuki2010, Bai2013, Fromang2013, Lesur2013}. It is likely that both the Hall effect and magnetic winds play a significant role in the overall evolution of protoplanetary discs, driving gas accretion at a much larger range of radii than the MRI \citep[e.g.][]{Bai2017}. Magnetic winds in particular could be shaping the inner disc structure along with the MRI \citep{Suzuki2016}. Nevertheless, the structure of the innermost regions of discs is still likely to be strongly affected by the MRI, and especially so the disc midplane where planets are expected to form. Therefore, while we do not consider non-viscous drivers of accretion in this work, the models presented here should still offer important insights for future work on the inner disc structure.

Finally, in this work we have also assumed that where the disc is not sufficiently ionized to drive the MRI, the viscosity parameter obtains a minimum, floor value ($\alpha_{\rm DZ}$). The value of this parameter determines whether dust will accumulate at the inner edge of the dead zone (i.e., at the pressure maximum; see Section \ref{sec:dust_accumulation}). An assumption of a fixed, non-zero $\alpha_{\rm DZ}$ in the MRI-dead zone is reasonable if such a viscosity can be driven by non-magnetic instabilities. The range of values explored in this paper covers the values observed in simulations of various hydrodynamic instabilities \citep[e.g.][]{Lesur2010, Nelson2013, Stoll2014}. However, outwards from the pressure maximum accretion may also be driven by propagation of waves from the adjacent MRI-active zone, and/or those outer regions may be heated by radial transport of heat from the MRI-active zone \citep[e.g.][]{Latter2012, Faure2014}. This would be contrary to our assumptions that the outer regions are heated by an uncorrelated source of viscosity and that the disc structure at different radii is uncorrelated except for the heating by stellar irradiation. In this case, the physics setting the location of the pressure maximum is more complex and the pressure maximum would likely occur at shorter orbital distances than predicted here. Ultimately though, at larger radii non-viscous drivers of accretion discussed above are likely to take over the evolution of the disc. If these are relevant at the outer edge of the MRI-active zone, the pressure maximum may still exist, provided that the gas still accretes faster in the inner region than in the outer. In this case, the disc midplane could be non-turbulent, allowing grain growth beyond the sizes predicted in Section \ref{sec:disc_dust_growth} and promoting accumulation of grains at the pressure maximum. Lastly, the inner edge of the MRI-dead zone is possibly unstable to formation of vortices \citep{Lyra2012, Faure2014}, which would invalidate our assumption of an azimuthally symmetric disc, but also possibly further promote accumulation of dust and planet formation.

\section{Summary} \label{sec:summary}
We have explored how the structure of the MRI-accreting inner regions of protoplanetary discs changes as a function of the dust-to-gas ratio, dust grain size, and other disc and stellar parameters. We have especially focused on the location of the gas pressure maximum arising at the boundary between the highly-viscous innermost region and the low-viscosity outer region. The existence and the location of the pressure maximum, and the disc structure in its vicinity, are key to the formation of the super-Earths inside the water ice line. 

At fixed dust parameters, the radius of the pressure maximum is directly related to the stellar mass $M_*$ and the gas accretion rate $\dot{M}$. This is because the stellar mass and the accretion rate determine the total viscous dissipation at a given radius, and thus the temperature and the ionization fraction at disc midplane. The radius of the pressure maximum is inversely related to the assumed viscosity parameter in the MRI-dead zone $\alpha_{\rm DZ}$. The location of the pressure maximum corresponds to a minimum in the viscosity parameter. Even though in our model there is an MRI-active layer at all radii, in the outer regions this is a (X-ray ionized) layer high above the disc midplane (in the vicinity of the pressure maximum), and the disc primarily accretes through the dense MRI-dead regions around the midplane. Therefore, the minimum viscosity parameter is close in value to the dead-zone $\alpha_{\rm DZ}$.

However, this picture of the highly-viscous innermost region and the low-viscosity outer region may change qualitatively for some disc and stellar parameters. Specifically, for low gas accretion rates ($\leq 10^{-9}$\,M$_\odot$\,yr$^{-1}$), we find that a steady-state solution could exist in which there is no gas pressure maximum. In these solutions the disc features low gas surface densities and high viscosity driven by X-ray ionization. Such solutions are more likely to exist for larger grain sizes (and, equivalently, at lower dust-to-gas ratios).

At fixed stellar and disc parameters, as long as the pressure maximum does exist, its location moves radially outwards as the dust grains grow to $a_{\rm max} \sim 10^{-2}$\,cm. Grain growth to still larger sizes results in the pressure maximum moving inwards, towards the star. This behaviour is primarily driven by the effects of dust opacities on the disc thermal structure.

We calculate the location of the pressure maximum for the case of dust growth being limited by turbulent fragmentation. For a Solar-mass star and gas accretion rates in the range $10^{-9} - 10^{-7}$\,M$_\odot$\,yr$^{-1}$, this always places the pressure maximum outwards of 0.1\,AU. In this fragmentation-limited regime, the radius of the pressure maximum depends very weakly on the dead-zone viscosity parameter, and it is most sensitive to the gas accretion rate. The pressure maximum may possibly not exist for a Solar-mass star and a gas accretion rate of $\leq 10^{-9}$\,M$_\odot$\,yr$^{-1}$, nor for a stellar mass of $0.1$\,M$_\odot$ for gas accretion rates $\leq 10^{-10}$\,M$_\odot$\,yr$^{-1}$, if the disc evolves into the high-viscosity X-ray ionized structure as soon as such structure can match the required disc accretion rate. This suggests that planet formation in the inner disc is more likely early in the disc lifetime.

The fragmentation-limited dust grain size and its Stokes number are most sensitive to the value of the viscosity (and turbulence) parameter at the pressure maximum. As noted above, this roughly equals the assumed value of the viscosity parameter in the MRI-dead zone ($\alpha_{\rm DZ}$). Therefore, whether the dust grains can become trapped in the pressure maximum is determined by this uncertain parameter. Dust trapping is likely for the lower end of plausible values ($\alpha_{\rm DZ}=10^{-5}$) and will not happen for the higher end ($\alpha_{\rm DZ}=10^{-3}$). 

Importantly, the pressure maximum does not move inwards for higher dust-to-gas ratios. That is, dust accumulation near the pressure maximum (and/or inwards of it) should result in an expansion of the dust-enriched region and/or dynamical evolution of the disc structure. However, time-dependent simulations are needed to further study the potential outcomes, and the viability of planetesimal formation in the inner disc.

\section*{Acknowledgements}
We thank the reviewer for their thoughtful comments. We thank Richard Booth, Thomas Haworth, Zhaohuan Zhu, Steven Desch, Neal Turner, Colin McNally and Henrik Latter for helpful discussions. MRJ acknowledges support from the President's PhD scholarship of the Imperial College London and the UK Science and Technology research Council (STFC) via the consolidated grant ST/S000623/1. JEO is supported by a Royal Society University Research Fellowship. This project has received funding from the European Research Council (ERC) under the European Union’s Horizon 2020 research and innovation programme (Grant agreement No. 853022, ERC-STG-2019 grant, PEVAP). JCT acknowledges support from NASA ATP grant Inside-Out Planet Formation (80NSSC19K0010).

\section*{Data Availability}
The data underlying this article will be shared on reasonable request to the corresponding author.




\bibliographystyle{mnras}
\bibliography{bibliography} 




\appendix


\bsp	
\label{lastpage}
\end{document}